\documentclass[9pt, aps, prd, nofootinbib, twocolumn]{revtex4-2}

\usepackage{amsmath}
\usepackage{amssymb}
\usepackage{bm}% bold math
\usepackage{braket}
\usepackage[usenames, dvipsnames]{color}
\usepackage{dcolumn}% Align table columns on decimal point
\usepackage{diagbox} 
\usepackage{epsfig}
\usepackage{slashed}
\usepackage{graphicx}% Include figure files
\usepackage{mathrsfs}
\usepackage{newtxmath, newtxtext}
\usepackage{subfigure}
\usepackage{mathrsfs}

 \usepackage[pagebackref=false, colorlinks=true]{hyperref}
\definecolor{reddish}{rgb}{0.7,0.2,0.0}  
\definecolor{blueish}{rgb}{0.1,0.1,1}
\hypersetup{
linkcolor=reddish,         % color of internal links
citecolor=blueish,           % color of links to bibliography
filecolor=blue,              % color of file links
urlcolor=magenta}       % color of the url

\def\ajp{Am. J. Phys.}           % American Journal of Physics
\def\araa{ARA\&A}                % Annual Review of Astron and Astrophys 
\def\apj{ApJ}                    % Astrophysical Journal 
\def\apjl{ApJL}                  % Astrophysical Journal, Letters 
\def\aap{A\&A}                   % Astronomy and Astrophysics 
\def\cqg{Class. Quantum Grav.}   % Classical and Quantum Gravity 
\def\epjc{Eur. Phys. J. C}       % European Physical Journal C
\def\mnras{MNRAS}                % Monthly Notices of the RAS 
 
\def\prd{Phys.~Rev.~D}           % Physical Review D 
\def\prl{Phys.~Rev.~Lett.}       % Physical Review Letters 

\def\sciadv{Sci. Adv.}

\begin{document}

\title[Static Photon Rings]{Prospects for Future Experimental Tests of Gravity with Black Hole Imaging: Spherical Symmetry}

\author{
Prashant Kocherlakota,$^{1,2,3}$
Luciano Rezzolla,$^{3,4,5}$ 
Rittick Roy,$^6$ and
Maciek Wielgus$^7$
}

\affiliation{
$^1$Black Hole Initiative at Harvard University, 20 Garden St., Cambridge, MA 02138, USA\\
$^2$Center for Astrophysics, Harvard \& Smithsonian, 60 Garden St., Cambridge, MA 02138, USA\\
$^3$Institut f{\"u}r Theoretische Physik, Goethe-Universit{\"a}t, Max-von-Laue-Str. 1, 60438 Frankfurt, Germany\\
$^4$School of Mathematics, Trinity College, Dublin 2, Ireland\\
$^5$Frankfurt Institute for Advanced Studies, Ruth-Moufang-Str. 1, 60438 Frankfurt, Germany\\
$^6$Anton Pannekoek Institute for Astronomy, University of Amsterdam, Science Park 904, 1098 XH, Amsterdam, The Netherlands\\
$^7$Max-Planck-Institut f{\"u}r Radioastronomie, Auf dem H{\"u}gel 69, D-53121 Bonn, Germany
}

\begin{abstract}
Astrophysical black holes (BHs) are universally expected to be described by the Kerr metric, a stationary, vacuum solution of general relativity (GR). Indeed, by imaging M87$^\star$ and Sgr A$^\star$ and measuring the size of their shadows, we have substantiated this hypothesis through successful null tests. Here we discuss the potential of upcoming improved imaging observations in constraining deviations of the spacetime geometry from that of a Schwarzschild BH (the nonspinning, vacuum GR solution), with a focus on the photon ring. The photon ring comprises a series of time-delayed, self-similarly nested higher-order images of the accretion flow, and is located close to the boundary of the shadow. In spherical spacetimes, these images are indexed by the number of half-loops executed around the BH by the photons that arrive in them. The delay time offers an independent shadow size estimate, enabling tests of shadow achromaticity, as predicted by GR. The image self-similarity relies on the lensing Lyapunov exponent, which is linked to photon orbit instability near the unstable circular orbit.  Notably, this critical exponent, specific to the spacetime, is sensitive to the $rr-$component of the metric, and also offers insights into curvature, beyond the capabilities of currently available shadow size measurements. The Lyapunov time, a characteristic instability timescale, provides yet another probe of metric and curvature. The ratio of the Lyapunov and the delay times also yields the lensing Lyapunov exponent, providing alternative measurement pathways. Remarkably, the width of the first-order image can also serve as a discriminator of the spacetime. Each of these observables, potentially accessible in the near future, offers spacetime constraints that are orthogonal to those of the shadow size, enabling precision tests of GR.
\end{abstract}

\keywords{astrophysical black holes, black hole imaging, null tests of general relativity, experimental tests of gravity, hotspots}

\maketitle

% ############################################################
% Main Text
% ############################################################

The Event Horizon Telescope (EHT) Collaboration has recently imaged the supermassive compact objects M87$^\star$ \cite{EHTC+2019a} and Sgr A$^\star$ \cite{EHTC+2022a}, adding to the mounting evidence indicating the ubiquitous existence of Kerr black holes (BHs), which are the spinning, vacuum BH solutions of general relativity (GR), at the centers of galaxies. Images of both EHT sources reveal a telltale dark region in the center that is surrounded by a bright emission ring, features that are typical in the synthetic images constructed from the simulations of accretion of hot, magnetized plasma onto Kerr BHs, which are used to model the astrophysical conditions of such objects \cite{Falcke+2000, Yuan+2014, EHTC+2019e, EHTC+2022d}. Optical transparency at 1.3$\mathrm{mm}$, the EHT observing wavelength for these sources, implies that the observed central intensity depression is best explained by the presence of a photon shell in the spacetime \cite{EHTC+2022f}, assured to exist in a typical BH spacetime \cite{Cunha+2020, Ghosh+2021}, and which casts a shadow on the observer's screen \cite{Bardeen1973}.

Since the photon shell and the shadow boundary curve are determined purely by the spacetime geometry (also the observer inclination angle for the latter), we realize immediately that approximate measurements of the shadow boundary curve, e.g., of its size, can be used to set up experimental tests of the spacetime geometry \cite{Johannsen+2010}. Indeed, the size of the observed bright emission ring in the image can be used to infer the shadow size of M87* \cite{EHTC+2019f, Psaltis+2020, Kocherlakota+2021} and of Sgr A* \cite{EHTC+2022f},%
\footnote{
The procedures to infer the shadow sizes of M87* and Sgr~A* differ from each other and thus cannot be directly compared currently. For Sgr~A*, see the discussion on $\alpha_1$-calibration in Sec. 3 of Ref. \cite{EHTC+2022f} for further details.} %
and the EHT finds that these are consistent with those of Kerr BHs of their respective masses \cite{EHTC+2019f, EHTC+2022f}. Together with gravitational-wave measurements involving stellar-mass BHs \cite{LSC+2019, LSC+2021a, LSC+2021b}, the EHT observations demonstrate the success of GR in describing the strong-field gravity near astrophysical BHs. 

Furthermore, the shadow size can be used to cleanly test the ``no-hair conjecture,'' which has been used to posit that all astrophysical BHs are Kerr BHs \cite{Johannsen+2010, Psaltis+2020, Volkel+2021, Kocherlakota+2022, EHTC+2022f}. For this reason, the Kerr BH metric is used almost exclusively when modeling astrophysical BHs. However, its interior geometry theoretically possesses several pathological features such as a Cauchy horizon, a spacetime singularity, as well as regions that permit closed timelike curves. Expanding our scope to consider phenomenological ``regular'' alternative BH spacetimes to the Kerr metric and looking for potential observable signatures in BH images can guide us towards resolutions of such pathologies \cite{Kumar+2020, Ghosh+2021b, Kocherlakota+2021, EHTC+2022f, Vagnozzi+2023}.

Moreover, horizonless compact objects can also possess photon shells, and can naturally also cast shadows \cite{Teo1998, Hioki2008, Chowdhury+2012, Sakai+2014, Cunha+2016, Grandclement2017, Shaikh2018, Shaikh+2019, Rahaman+2021, Ghosh+2021, Shaikh+2021, Solanki+2022}. Therefore, observations of the shadow cast by an astrophysical compact object allow us to potentially distinguish between -- and possibly rule out -- different types of BHs, naked singularities, wormholes, gravastars, and other exotic objects that may be \textit{a priori} allowed models \cite{Volkel+2021, Kocherlakota+2021, EHTC+2022f, Vagnozzi+2023}. These could also enable experimental tests of the weak cosmic censorship conjecture \cite{Wald1997}. 

Finally, since these BH and non-BH models arise as solutions in different theories of classical gravity, theories with new fields, or theories with alternative field couplings to gravity, such constraints can be used to distinguish between underlying theories, as emphasized in previous work \cite{Kumar+2020b, Afrin+2021, Volkel+2021, Kocherlakota+2021, EHTC+2022f, Kumar+2022}. 

Present operating angular resolution prevents access to finer details such as the ellipticity of the shadow boundary, which encodes additional information about the spacetime, such as the BH spin. Future experiments at higher angular resolutions and/or flux-sensitivities may be able to overcome some limitations, allowing for unprecedented experimental tests of the Kerr metric \cite{Gralla+2020b} as well as sharper constraints on alternative spacetimes \cite{Wielgus2021, Ayzenberg2022, Broderick+2023, Staelens+2023, Salehi+2023}. Indeed, methods have been proposed to detect the ``inner shadow'' \cite{Chael+2021}, the lensing Lyapunov exponent \cite{Johnson+2020, Hadar+2021}, the delay time \cite{Gralla+2020a, Wong2021, Hadar+2021}, and the Lyapunov time \cite{Ames+1968, Cardoso+2009, Cardoso+2021}. Further work analyzing the feasibility of obtaining such measurements can be found in Refs. \cite{Paugnat+2022, Vincent+2022, Kocherlakota+2024a}.

In this demonstrative study, restricting to general spherically-symmetric and static spacetimes, our first goal is to highlight the relationships between the various aforementioned critical parameters (cf. Ref. \cite{Kocherlakota+2024a}). These, we hope, will provide new avenues to independently determine some parameters. For example, the ratio of the characteristic delay time and the shadow size is always equal to $\pi$. Thus, a measurement of the delay time from light curves of flaring events, obtained at relatively lower angular resolution (possible also at different frequencies), could provide qualitatively similar spacetime information as the current high-resolution EHT measurements. Furthermore, while some parameters may be more challenging to measure (such as the lensing Lyapunov exponent), these relationships can assure us that equivalent information about the spacetime geometry is encoded in others that may be easier to quantify. We find, for example, that a simple ratio of the delay and the Lyapunov time equals the lensing Lyapunov exponent. Thus, a single additional measurement of the Lyapunov time, from future observations, combined with a measurement of the shadow size, which is already possible \cite{Psaltis+2020, Kocherlakota+2021, EHTC+2022f}, can lead to an inference of the lensing Lyapunov exponent. We note that all of these critical parameters (in addition to $\delta_0$, which is nontrivial in spinning spacetimes) have been obtained also for the spinning Kerr BH spacetime \cite{Gralla+2020a}. The relations between the Kerr critical parameters were examined in Ref. \cite{Hadar+2022}, with which our findings in general nonspinning spacetimes are consistent. Combined, both these findings indicate that such relations may remain true for a general class of spinning non-Kerr spacetimes (cf. Ref. \cite{Johannsen2013}).

Our second goal here is to understand the implications of measurements of the critical exponents for measurements of the underlying spacetime metric. In particular, we examine here constraints on spherically-symmetric deviations from the Schwarzschild BH metric, which is the nonspinning vacuum solution of GR. We achieve this by employing the Rezzolla-Zhidenko \cite{Rezzolla+2014} metric-deviation parameter spaces, which have previously been used to explore the impact of currently available EHT shadow-size measurements \cite{Psaltis+2020, Volkel+2021, Bauer+2022, Kocherlakota+2022, EHTC+2022f}. We will revisit below how while these current measurements impose nontrivial constraints on these parameter spaces, there remain unconstrained directions \cite{Psaltis+2020, Volkel+2021, Kocherlakota+2022}. We demonstrate here how these unconstrained directions become bounded by an additional measurement of either the lensing Lyapunov exponent or the Lyapunov time. Furthermore, these additional measurements also access new aspects of the spacetime metric (its $rr-$component as well as its curvature) that current shadow-size measurements are completely oblivious to. Finally, we find indications that the width of the first-order image can also be used conditionally to infer information about the spacetime geometry.

In conclusion, we anticipate that future black hole imaging measurements will yield stronger tests of the no-hair conjecture and help build confidence in our theoretical understanding of the properties of astrophysical black holes, as well as provide novel null tests of GR.

One major limitation of our work is ignoring the impact of the black hole spin. While we expect our findings to carry over to the metric-deviation parameter spaces describing spinning black holes, a detailed exploration of the same is nonetheless crucial.

\section{Spacetime Critical Parameters}
\label{sec:SecI_Scaling_Relations}

In this section, we review the ``critical parameters'' of the spacetime, the primary observables of interest here. These include the shadow size $\eta_{\mathrm{PS}}$, the lensing Lyapunov exponent $\gamma_{\mathrm{PS}}$, the Lyapunov time $t_{\ell; \mathrm{PS}}$, and the delay time $t_{d; \mathrm{PS}}$. For details regarding the derivations for the results presented here, we direct the reader to see our companion paper \cite{Kocherlakota+2024a} as well as earlier pioneering work (see, e.g., Refs. \cite{Cardoso+2009, Bozza+2007, Gralla+2019, Gralla+2020a}). 

The line element of an arbitrary static and spherically-symmetric spacetime can be expressed in spherical-polar coordinates $x^\alpha = (t, r, \vartheta, \varphi)$ as
\begin{equation} \label{eq:Static_Spacetime} 
\mathrm{d}s^2 = \hat{\mathscr{g}}_{\mu\nu}\mathrm{d}x^\alpha\mathrm{d}x^\beta 
= -f~\mathrm{d}t^2 + \frac{g}{f}~\mathrm{d}r^2 + R^2~\mathrm{d}\Omega_2^2\,,
\end{equation}
where the metric functions $f, g$, and $R$ are functions of $r$ alone, and $\mathrm{d}\Omega_2^2 = \mathrm{d}\vartheta^2 + \sin^2{\vartheta}~\mathrm{d}\varphi^2$ is the standard line element on a unit 2-sphere. We will assume reasonably that $g>0$ everywhere and that $R > 0$ except at the center ($R=0$). The metric above describes a BH spacetime if $f(r)$ admits real, positive zeroes (with $R>0$), the largest of which locates the event horizon, which we denote by $r_{\mathrm{H}}$. For a Schwarzschild BH of mass $M$, we have $f(r) = 1 - 2M/r, g(r) = 1, R(r) = r$ and $r_{\mathrm{H}} = 2M$.

We adopt here the convention of defining, without loss of generality (due to spherical symmetry), the inclination of the observer to be zero ($\vartheta_{\mathrm{o}} = 0$). Due to the planarity of geodesic orbits in spherically-symmetric spacetimes, this has the profoundly simplifying consequence that \textit{all} photons arriving at this observer move only on meridional planes ($\mathrm{d}\varphi = 0$) through space. If we denote the four-velocity along an arbitrary \textit{meridional} photon orbit by $k^\mu := \dot{x}^\mu$, where the overdot represents a derivative w.r.t. the affine parameter $\lambda$ along it, then we can introduce the photon orbit radial $\mathscr{R}$ and polar $\Theta$ effective potentials as
\begin{align} \label{eq:Null_Geodesic_Effective_Potential}
\mathscr{R}(\eta, r) :=&\ (\dot{r}/E)^2 = g^{-1}\left[1-\eta^2 f R^{-2}\right]\,, \\
\Theta(\eta, r) :=&\ (\dot{\vartheta}/E)^2 = \eta^2/R^4\,.
\end{align}
Here $E$ is the energy of the photon and $\eta := |p_\vartheta|/E$ is the ratio of its total angular momentum and energy, both conserved quantities. The latter is also the (apparent) impact parameter of the photon, i.e., it is the radius at which it appears on the image plane \cite{Bardeen1973}. 

Due to the strong gravity near ultracompact objects, it is generically possible for photons to move on circular orbits \cite{Cunha+2020, Ghosh+2021}. For such photons, $\dot{r} = \ddot{r} = 0$, or equivalently, $\mathscr{R} = \partial_r\mathscr{R} = 0$. These yield the photon sphere radius $r_{\mathrm{PS}}$ and the critical impact parameter $\eta_{\mathrm{PS}}$ (this is the shadow radius) as
\begin{equation} \label{eq: Photon_Sphere_Shadow_Boundary}
(\partial_r f)/f - 2(\partial_r R)/R = 0\,;\ 
\eta_{\mathrm{PS}} := R_{\mathrm{PS}}/\sqrt{f_{\mathrm{PS}}}\,.
\end{equation}
Here the subscript ``PS'' for a metric function indicates that it is evaluated at $r=r_{\mathrm{PS}}$, e.g., $R_{\mathrm{PS}} := R(r_{\mathrm{PS}})$. We assume the first equation above admits a single root outside the event horizon (however, cf. Refs. \cite{Wielgus+2020, Gan+2021, Guo+2022}). For the Schwarzschild BH spacetime, $r_{\mathrm{PS}} = 3M$ and $\eta_{\mathrm{PS}} = \sqrt{27}M$.

The total angular deflection $\slashed{\Delta}\vartheta^\pm$ experienced by a meridional photon emitted from a spatial location $(r_{\mathrm{e}}, \vartheta_{\mathrm{e}})$ with impact parameter $\eta$ is given heuristically as $\slashed{\Delta}\vartheta^\pm(\eta, r_{\mathrm{e}}) = \pm\fint_{r_{\mathrm{e}}}^\infty\sqrt{\Theta/\mathscr{R}}~\mathrm{d}r$, where the slash represents a path-dependent integral. Due to the absence of nontrivial ``polar turning points'' (where $\dot{\vartheta} = 0$) in a spherically-symmetric spacetime, the sense of rotation of a photon about the center ($r=0$) remains invariant along its orbit. Therefore, since photons emitted from the same spatial location but with opposite polar velocities (different signs of $\dot{\vartheta}$) appear at diametrically opposite points on the image plane, we will use superscripts ``$\pm$'' to keep track of this aspect.

It is clear to see from the above that for a circular photon orbit the total deflection angle $\slashed{\Delta}\vartheta^\pm$ diverges because the radial potential vanishes. If we introduce the bulk $\bar{r}$ and boundary $\bar{\eta}$ conformal radii respectively as 
\begin{align}
\bar{r} := r/r_{\mathrm{PS}} - 1\,,\ \ 
\bar{\eta} := \eta/\eta_{\mathrm{PS}} - 1\,, 
\end{align}
it follows that photons which become close to the photon sphere ($|\bar{r}| \ll 1$) somewhere along their orbit and whose impact parameters are close to the critical one ($|\bar{\eta}| \ll 1$) must also experience strong gravitational lensing (since $|\mathscr{R}| \ll 1$). Indeed, the photon ring is identified as the region on the image plane where the total deflection angle diverges (logarithmically) as $\slashed{\Delta}\vartheta^\pm \propto \ln|\bar{\eta}|$ \cite{Johnson+2020}. 

More precisely, we can write the following scaling relations for the total deflection angle and total orbital time for photons that appear in the photon ring, in general static and spherically-symmetric spacetimes, as \cite{Kocherlakota+2024a}
\begin{align} \label{eq:Scaling_Relations}
\begin{alignedat}{3}
\slashed{\Delta}\vartheta^\pm \approx 
&\ \mp \frac{\pi}{\gamma_{\mathrm{PS}}}\ln{|\bar{\eta}|}\,,\ 
&& \quad\quad 
\gamma_{\mathrm{PS}} :=\ \frac{\pi R^2_{\mathrm{PS}}}{\eta_{\mathrm{PS}}}\hat{\kappa}_{\mathrm{PS}}\,; \\
%$###########################
\slashed{\Delta}t^\pm \approx 
&\  -t_{\ell; \mathrm{PS}}\ln{|\bar{\eta}|}\,,
&& \quad\quad
t_{\ell; \mathrm{PS}} :=\ \frac{1}{f_{\mathrm{PS}}\hat{\kappa}_{\mathrm{PS}}}\,.
\end{alignedat}
\end{align}
The constants introduced above, $\gamma_{\mathrm{PS}}$ and $t_{\ell; \mathrm{PS}}$, are the lensing Lyapunov exponent and the Lyapunov time respectively. We will see presently how they impact various observables. Finally, $\hat{\kappa}_{\mathrm{PS}}$ is a constant that is given as \cite{Kocherlakota+2024a}
\begin{align} \label{eq:Phase_Space_Lyapunov_Exponent}
\hat{\kappa}^2_{\mathrm{PS}} := -\frac{1}{2g_{\mathrm{PS}}}\left(\frac{\partial_r^2f_{\mathrm{PS}}}{f_{\mathrm{PS}}} - \frac{\partial_r^2R^2_{\mathrm{PS}}}{R^2_{\mathrm{PS}}}\right)\,.
\end{align}
Clearly, this plays a key role in determining various fundamental quantities and can be understood as follows. The radial evolution of a photon with critical impact parameter that is initially present close to the photon sphere, $|\bar{r}(\lambda = 0)| \ll 1$ is given as $\bar{r}(\lambda) = \bar{r}(0)\exp{\left[\pm_r E \hat{\kappa}_{\mathrm{PS}}\lambda\right]},$ where $\pm_r$ denotes the initial sign of the photon's radial velocity. Therefore, $\hat{\kappa}_{\mathrm{PS}}$ is the (fundamental) phase space Lyapunov exponent that governs the radial instability of photon orbits at the photon sphere, and is related to a certain component of the Riemann tensor (cf. Refs. \cite{Blaga+2023, Kocherlakota+2024a}). This is to be expected since the previous equation is a geodesic deviation equation for null geodesics and is also related to the Raychaudhuri equation for the meridional null congruence. For the Schwarzschild BH spacetime, $\hat{\kappa}_{\mathrm{PS}} = 1/(\sqrt{3}M)$.

We can rewrite this radial evolution in terms of the coordinate time as $\bar{r}(t) = \bar{r}(0)\exp{[\pm_r t/t_{\ell; \mathrm{PS}}]}$  (see also Ref. \cite{Cardoso+2009}). Thus, the Lyapunov time $t_{\ell; \mathrm{PS}}$ is the characteristic instability timescale for photons located radially-close to the photon sphere. That is, it is the time, as measured by an asymptotic ($r\rightarrow\infty$) static observer ($u \propto \partial_t$), for the radial coordinate between photon orbits, close to the photon sphere, to increase by a factor of $\mathrm{e} \approx 2.72$. For the Schwarzschild BH spacetime, $t_{\ell; \mathrm{PS}} = \sqrt{27}M$. 

Recently, in Ref. \cite{Cardoso+2021} (see also Ref. \cite{Ames+1968}), it was shown that this time scale plays an important role in determining the late-time characteristics of the observed luminosity evolution (light curve) of a star falling into a BH. Thus, the Lyapunov time can, in principle, be measured. It remains to be seen whether this can also be obtained from the late-time behavior of a light curve corresponding to a flaring event associated with Sgr A$^\star$ (see also the delay time below). Alternatively, observing a gas cloud falling into Sgr A$^\star$ \cite{Moriyama+2019} for an extended time may be suitable for such a measurement to be made. 

The lensing Lyapunov exponent can be understood as follows. There exist photons emitted from the same initial spatial location and captured by an observer at the same final spatial location that have orbits differing in impact parameter, total angular deflection, elapsed affine parameter (or total path-length), and elapsed coordinate time. These are all photons for which we can write $\slashed{\Delta}\vartheta^\pm = (\vartheta_{\mathrm{o}} - \vartheta_{\mathrm{e}}) \mod{2\pi} = -\vartheta_{\mathrm{e}}\mod{2\pi}$ (see, e.g., Refs. \cite{Gralla+2019, Cunha+2020, Wei2020, Xu+2023}). This is equivalently expressed as \cite{Kocherlakota+2024a}
\begin{equation} \label{eq:Image_Order_Definition}
\slashed{\Delta}\vartheta^\pm_n 
= \pi/2 - \vartheta_{\mathrm{e}} + (-1)^{n+1}(2n+1)\pi/2\,,
\end{equation}
where $n$ is the order of the photon or image. The absolute total angular deflection for this class of orbits increases with image order, taking values in $\{(0, \pi), (\pi, 2\pi), (2\pi, 3\pi), (3\pi, 4\pi), \cdots \}$ respectively. The case when the source is located along the observer axis must be treated separately.%
\footnote{A point source present at $\vartheta_{\mathrm{e}} = 0$ or $\pi$ does not form discrete images. Instead, images are entire rings, called Einstein rings or ``critical curves'' (cf. Ref. \cite{Virbhadra+2000}). Such source locations are called caustics, which are defined as locations in the past light cone of the observer that have divergent magnifications. Critical curves are the maps of caustics on the image plane \cite{Virbhadra+2000, Frittelli+2000}. Thus, it becomes clear that the $n=\infty$ critical curve coincides with the shadow boundary curve \cite{Gralla+2019}.} %
Remembering that the sign ($\pm$) denotes the sign of the initial polar velocity, we see from this equation that even-order photons ($n=0, 2, \cdots$) come with a negative sign and vice versa.

With the expression for $\slashed{\Delta}\vartheta^\pm_n$ in hand \eqref{eq:Image_Order_Definition}, we can use the angular deflection scaling relation \eqref{eq:Scaling_Relations} to obtain the relation between the image radii of consecutive orders for an arbitrary point source location as
\begin{equation} \label{eq:n_n+1_Radius_Scaling}
\frac{\bar{\eta}_{n+1}}{\bar{\eta}_n} \approx \mathrm{e}^{-\gamma_{\mathrm{PS}}}\cdot\mathrm{e}^{\pm\gamma_{\mathrm{PS}}(2\vartheta_{\mathrm{e}}/\pi-1)}\,. 
\end{equation}
In this equation, and similar equations below (\ref{eq:Width_Flux_Scaling}, \ref{eq:Time_Delay}), the upper sign is picked if $n$ is even and $(n+1)$ is odd, and vice versa.

Scaling relations that relate the characteristics of \textit{consecutive} order images have been obtained for equatorial sources of emission and/or when viewed by an observer at the pole \cite{Johnson+2020, Gralla+2020a}. The general equation above, for arbitrary sources and observers in spherical spacetimes, shows that the ``equatorial'' ($\vartheta_{\mathrm{e}} = \pi/2$) scaling relations receive order-unity corrections for nonequatorial emitters.

In astrophysical accreting systems such as M87$^\star$ and Sgr A$^\star$, emission is sourced by both the accretion flow (``disk'') and an outflow (``jet'' or ``wind''). Since the scale height of the disk is rather small ($h/r \lesssim 0.4$; cf. Ref. \cite{Narayan+2022}), modeling its emission as being primarily equatorial is a good approximation. Emission from the jet (sheath) is primarily sourced from off the equator and can contribute a significant amount of flux density on the image plane. Thus, the extended relations above \eqref{eq:n_n+1_Radius_Scaling} could play a useful role in establishing expectations regarding inferences of the lensing Lyapunov exponent from observations (cf. Ref. \cite{Kocherlakota+2024a}).

As discussed there, for an extended conical surface of emission, $\vartheta=\vartheta_{\mathrm{e}}\neq\pi/2$, viewed face-on, it follows from eq. \ref{eq:n_n+1_Radius_Scaling} that the image radii $\eta_n$, widths $w_n$, and their flux densities $F_n$ of consecutive order images approximately satisfy
\begin{align} \label{eq:Width_Flux_Scaling}
\frac{\eta_{n+1} - \eta_{\mathrm{PS}}}{\eta_n - \eta_{\mathrm{PS}}} \approx
\frac{w_{n+1}}{w_n} \approx 
\frac{F_{\nu; n+1}}{F_{\nu; n}} \approx
\mathrm{e}^{-\gamma_{\mathrm{PS}}}\cdot\mathrm{e}^{\pm\gamma_{\mathrm{PS}}(2\vartheta_{\mathrm{e}}/\pi-1)}\,. 
\end{align}
For emission from the equatorial plane, $\vartheta=\vartheta_{\mathrm{e}} = \pi/2$, the right hand side reduces simply to $\mathrm{e}^{-\gamma_{\mathrm{PS}}}$ \cite{Johnson+2020}. Thus, the lensing Lyapunov exponent controls the self-similar scaling of higher-order images, and it has recently been argued that this could be measured from future black hole imaging measurements \cite{Johnson+2020}. For the Schwarzschild BH spacetime, $\gamma_{\mathrm{PS}} = \pi$ \cite{Ohanian1987, Bozza+2007, Gralla+2019}.

Furthermore, with eq. \ref{eq:n_n+1_Radius_Scaling} and the orbital time scaling relation \eqref{eq:Scaling_Relations}, we can obtain the time delay $\Delta t_n^{\pm} := \slashed{\Delta}t_{n+1}^{\mp} - \slashed{\Delta}t_n^{\pm}$ between consecutive order images as 
\begin{align} \label{eq:Time_Delay}
\Delta t_n \approx
\pi\eta_{\mathrm{PS}}\left[1 \mp \left(\frac{2\vartheta_{\mathrm{e}}}{\pi} - 1\right)\right] := t_{d; \mathrm{PS}}\left[1 \mp \left(\frac{2\vartheta_{\mathrm{e}}}{\pi} - 1\right)\right]\,.
\end{align}
In the above, we have finally introduced the characteristic delay time, $t_{d; \mathrm{PS}}$, which is an approximate measure of the time elapsed between the appearance of consecutive order images on the image plane. For a Schwarzschild BH, $t_{d; \mathrm{PS}} = \pi\sqrt{27}M$ \cite{Gralla+2019, Gralla+2020a}. The delay time is simply the half-orbital time of a photon moving on a circular meridional orbit,
\begin{equation}
\frac{t_{orb; \mathrm{PS}}}{2} = \frac{\pi}{\Omega_{\mathrm{PS}}} = \pi\eta_{\mathrm{PS}}\,,
\end{equation}
where $\Omega_{\mathrm{PS}} := 1/\eta_{\mathrm{PS}}$ is its angular velocity. This is a remarkable result: A clean detection of the time delay between higher-order images can yield an independent estimate of the shadow size $\eta_{\mathrm{PS}}$ in spherically-symmetric spacetimes. Since such a measurement is independent, in GR, of the frequency at which these observations are conducted, multifrequency observations of flaring events can potentially be used to set up null tests of the achromaticity of the BH shadow.

Compact flux eruption events, or flaring events, associated with Sgr A$^\star$ are observed across a multitude of wavelengths \cite{Eckart+2006, Gravity+2018}. Such sources of compact flux have been modeled in practice using hotspots (compact blobs of emission), moving in the BH equatorial plane, and it has been suggested that the time delay between its zeroth and first-order images can be measured from observations \cite{Tiede+2020, Ball+2021, Hadar+2021, Wong2021}. Another alternative to measuring this delay time comes from measuring the light curve of a gas cloud that is falling into a black hole \cite{Moriyama+2019}. Proposals to use such measurements to infer the spin of Sgr A$^\star$ \cite{Wong2021, Moriyama+2019} as well as to obtain information about the spacetime geometry \cite{Sahu+2013} have also been forwarded.

Another promising avenue for detecting higher-order images and measuring the delay time and the lensing Lyapunov exponent could be by constructing autocorrelations either of the light curve or of the intensity fluctuations in high-resolution movies of black holes, as described in Ref. \cite{Hadar+2021} and explored in Ref. \cite{Chen+2023b}.

We also point out that Ref. \cite{Cardoso+2009} (see also \cite{Stefanov+2010, Chen+2023a}) establishes a concrete connection between the shadow size and the Lyapunov time on the one hand and the quasinormal mode frequencies on the other, in arbitrary spherically-symmetric and static spacetimes,
\begin{equation}
\omega_{\mathrm{QNM}} = l\left(\frac{1}{\eta_{\mathrm{PS}}}\right) - \mathrm{i}\left(n+\frac{1}{2}\right)\left(\frac{1}{t_{\ell; \mathrm{PS}}}\right)\,,
\end{equation}
where $l$ and $n$ are the angular momentum and the overtone numbers of the quasinormal mode perturbation.

Finally, we note that a combined measurement of the Lyapunov and delay time yields an alternative and independent estimate of the lensing Lyapunov exponent,
\begin{equation}
\frac{t_{\ell; \mathrm{PS}}}{t_{d; \mathrm{PS}}} = \frac{\pi\eta_{\mathrm{PS}}/\gamma_{\mathrm{PS}}}{\pi\eta_{\mathrm{PS}}} = \frac{1}{\gamma_{\mathrm{PS}}} \,.
\end{equation}
This relationship has been shown to hold in the Kerr spacetime as well (cf. eq. 3.40 of Ref. \cite{Hadar+2022}). Together, these results provide a firm basis for finding similar ones in general axisymmetric spacetimes (cf., e.g., Ref. \cite{Salehi+2023}), which may play a vital role in developing methods for inferring harder-to-measure critical parameters (e.g., lensing Lyapunov exponent) from (relatively) easier-to-measure ones.

\section{Constraints on Spacetime from Critical Parameters}

The black hole no-hair conjecture has been used to posit that astrophysical BHs (not in dynamical scenarios like mergers) are described by just two numbers-- their mass $M$ and intrinsic angular momentum $J$. That is, all multipoles of the gravitational field \cite{Geroch1970} are determined by these numbers. A restricted empirical test of this conjecture becomes possible by first constructing parameter spaces that measure spherically-symmetric ($J=0$) deviations from a Schwarzschild BH and then examining the constraints induced by measurements such as the EHT measurements of the shadow sizes \cite{Johannsen+2010}. Indeed, this approach has provided nontrivial null tests of the conjecture \cite{Psaltis+2020, EHTC+2022f}. 

However, significant degeneracies remain \cite{Psaltis+2020, Volkel+2021, Kocherlakota+2022}, and, indeed, the spacetime metric of M87$^\star$ or Sgr A$^\star$ can differ by an arbitrarily large amount from that of a Schwarzschild BH in certain parametric directions. We will show below by using the Rezzolla-Zhidenko (RZ; \cite{Rezzolla+2014}) parametrization framework that a single additional measurement of the lensing exponent can render such presently unconstrained regions compact, thus making no-hair tests significantly more potent. Our analysis can easily be extended to other popular parametrization frameworks \cite{Johannsen2013, Vigeland+2011} to yield similar findings.

Several well-known static BH spacetimes, that arise as solutions in distinct alternative theories of gravity (and fields), have been approximated to very high accuracy using a small number ($\lesssim\!\!11$) of RZ metric deviation, or expansion, parameters \cite{Kocherlakota+2020}; This is possible since the RZ framework exploits the fantastic convergence properties of Pad{\'e} approximants. The ambit of the RZ framework has also been extended to arbitrary static spacetimes (including non-BHs) there. Furthermore, it has also been demonstrated that when using it to approximate observables, such as the shadow size, of known solutions, the accuracy required to enable comparisons against EHT measurements can be achieved with even fewer ($\lesssim\!\!3$) RZ parameters \cite{Konoplya+2020}. 

The original RZ metric functions, $N^2$ and $B^2$ \cite{Rezzolla+2014}, are related to the ones used here simply as, $N^2(r) = f(r)$ and $B^2(r) = g(r)$. Furthermore, the RZ metric sets $R(r) = r$. Here, since this suffices for our purposes, in addition to a parameter $\epsilon$, which exclusively controls the size of the event horizon, we will consider only the first few leading order parameters of the RZ metric $\{a_0, a_1, b_0, b_1\}$. Here we will only consider BHs of the same (Arnowitt-Deser-Misner; Ref. \cite{Arnowitt+2008}) mass $M$. As we shall see below, $a_0$ and $b_0$ control the asymptotics of the spacetime whereas $a_1$ and $b_1$ control the near-horizon geometry. The metric functions for this RZ family are then given explicitly as,
\begin{align} 
\label{eq:f_RZ_eps_a0_a1}
f(r) =&\ 1 - \frac{2M}{r} + \frac{4a_0}{(1+\epsilon)^2}\frac{M^2}{r^2} \\
& + \frac{8(\epsilon - a_0 + a_1)}{(1 + \epsilon)^3}\frac{M^3}{r^3}  -\frac{16 a_1}{(1+\epsilon)^4}\frac{M^4}{r^4}\,, \nonumber \\
\label{eq:g_RZ_b0_b1}
g(r) =&\ \left[1 + \frac{2b_0}{r} + \frac{4b_1}{r^2}\right]^2\,.
\end{align}
The location of the outermost horizon $r = r_{\mathrm{H}}$ is defined to be at \cite{Rezzolla+2014},
\begin{equation} \label{eq:RZ_Horizon}
r_{\mathrm{H}} := \frac{2M}{1+\epsilon}\,.
\end{equation}
Clearly, we will require that $\epsilon > -1$ for BH spacetimes. We emphasize that requiring the largest root of $f(r)$ be located at $r=r_{\mathrm{H}}$ automatically imposes non-trivial constraints on the RZ metric deviation parameter space. The ranges of the theoretically permissible RZ metric deviation parameters depend, in general, on the family of RZ metric in use. For three of the families of RZ metric that we will use here ($b_i = 0$), these constraints can be found in Table 2 of Ref. \cite{Kocherlakota+2022}. For the last family we use here ($a_i = 0$), the condition that $g(r)$ be nonvanishing everywhere ($r>0$) imposes the condition that $b_1 > b_0^2/4$. This is necessary simply to ensure that the proper volume of space inside a finite coordinate radius $r$ is nonzero everywhere. The boundary demarcating the permissible and impermissible regions is shown in all panels as a red line, with the latter shown as white regions. We note that these spacetimes can contain strong curvature singularities at their centers $r=0$ but are regular everywhere else \cite{Kocherlakota+2022}. 

\begin{figure*}
\centering
\includegraphics[width=1.9\columnwidth]{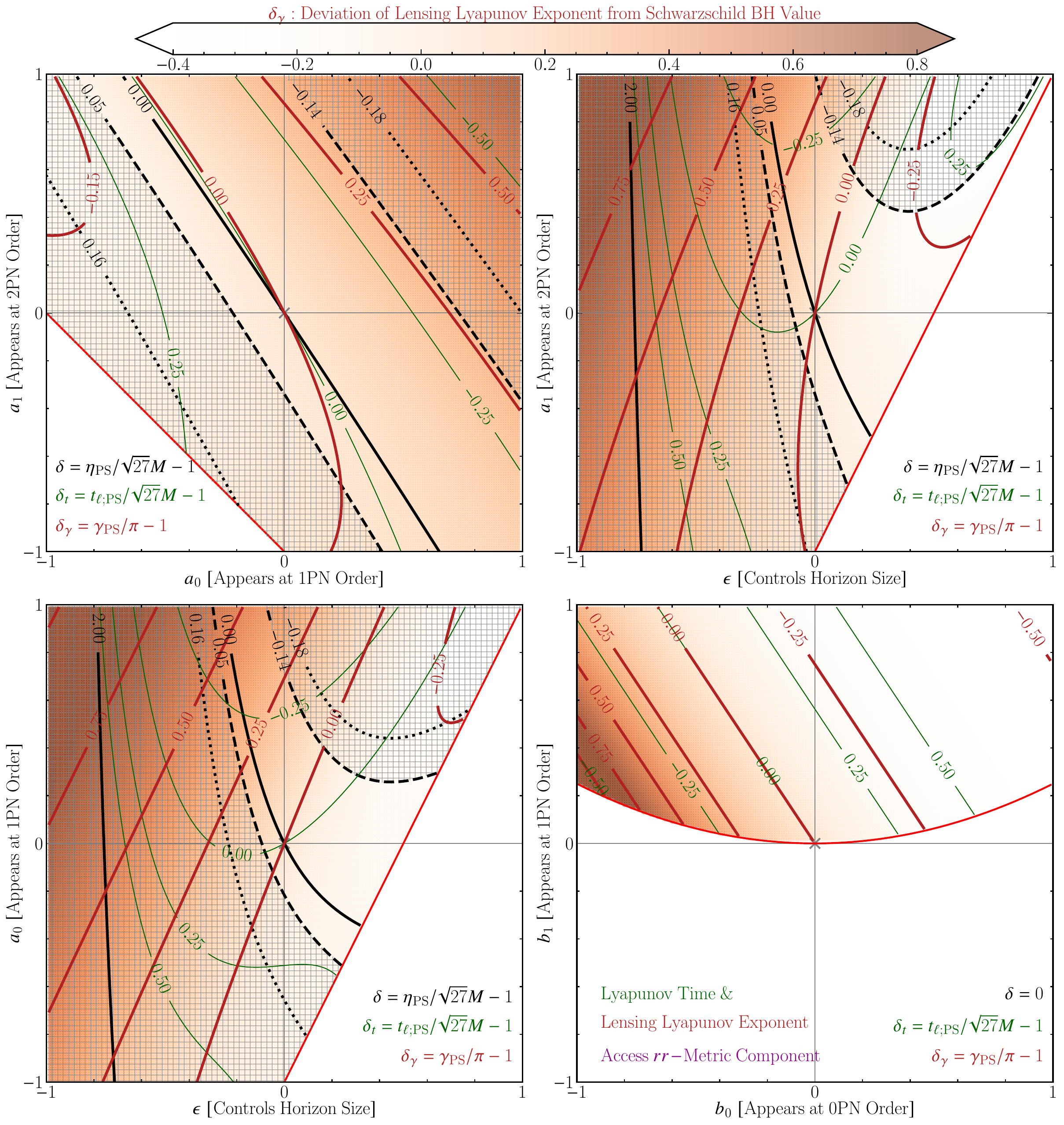}
\caption{\textit{Variation in characteristic spacetime observables with metric deviation parameters.} We show here the fractional deviations in the shadow size ($\eta_{\mathrm{PS}}$), the Lyapunov time ($t_{\ell; \mathrm{PS}}$), and the lensing Lyapunov exponent ($\gamma_{\mathrm{PS}}$) from their Schwarzschild values in black, green, and red lines respectively, for four different families of (RZ) black holes. All of these BHs have the same mass $M$, and the Schwarzschild BH is located at $(0, 0)$ in each panel. Each of these observables involves a different combination of the metric functions and their derivatives. Their respective isocontours intersect at unique locations generically in these BH metric deviation parameter spaces. This demonstrates, quite strikingly, how combining measurements of these observables would yield stringent and precision tests of general relativity in the strong-field regime. Each of these BH parameter spaces samples a qualitatively different type of metric deviation from the Schwarzschild spacetime: $\epsilon$ and $a_i$ control the $tt-$component of the metric whereas $b_i$ control its $rr-$component. The 2017 EHT M87$^\star$ and Sgr A$^\star$ shadow size ($1\sigma$) bounds are shown in dotted and dashed lines respectively (hatches indicate the region ruled out by the latter).}
\label{fig:Fig1_RZ_Pure_Metric_Observables}
\end{figure*}

From the above, it is evident that of all RZ BHs with the same mass $M$, only those for which $\epsilon = 0$ have the same horizon size as the Schwarzschild BH. 
It is also clear from eqs. \ref{eq:f_RZ_eps_a0_a1} and \ref{eq:g_RZ_b0_b1} that the first post-Newtonian (PN) coefficients ($1/r$-terms) are determined by $\epsilon, a_0,$ and $b_0$, whereas, for this class of RZ metrics, the higher-order RZ parameters, $a_1$ and $b_1$, control higher-PN coefficients. Moreover, the parametrized post-Newtonian (PPN) parameters, $\beta_{\mathrm{PPN}}$ and $\gamma_{\mathrm{PPN}}$, are given by particular combinations of these metric deviation parameters \cite{Rezzolla+2014}. In particular, PPN constraints obtained by solar system measurements \cite{Will2014} can be translated into constraints on combinations of the lowest-order RZ parameters as $|2b_0/(1+\epsilon)| \lesssim 2.3 \times 10^{-5}$ and $|2a_0/(1+\epsilon)^2+2b_0/(1+\epsilon)| \lesssim 2.3\times 10^{-4}$ \cite{Rezzolla+2014}. Therefore, finding similar constraints on $a_0$ and $b_0$ via black hole imaging measurements in the strong gravity near supermassive compact objects can help us compare the strength of obtained constraints across several magnitudes in gravitational-field strength \cite{EHTC+2022f}, and test the validity of the Birkhoff theorem (see, e.g., Ref. \cite{Wald1984}).

When the metric describing a spherically-symmetric and static spacetime is written in areal-polar coordinates, $R(r) = r$, as in the RZ metric, the locations of the horizon, photon sphere, and the innermost stable circular orbit (ISCO; This is the timelike Keplerian orbit that is closest to the compact object) are set by the $tt-$component of the metric alone (see, e.g., Ref. \cite{Kocherlakota+2020}). Fig. \ref{fig:FigA1_RZ_Pure_Metric_Characteristics} in the Appendix shows the variation of these quantities for all the RZ BH models we consider below.

We turn now in Fig. \ref{fig:Fig1_RZ_Pure_Metric_Observables} to the variation in the purely metric-dependent \textit{observables} with varying spacetime geometry. We define the three deviation parameters that we need as follows,
\begin{align} \label{eq:Deviation_Parameters}
\begin{alignedat}{2}
\delta &=\ \frac{d_{\mathrm{sh}}}{d_{\mathrm{sh; Schw}}} - 1 
&&=\ \frac{\eta_{\mathrm{PS}}}{\sqrt{27}M} - 1\,, \\
\delta_t &=\ \frac{t_{\ell; \mathrm{PS}}}{t_{\ell; \mathrm{ps; Schw}}} - 1 
&&=\ \frac{t_{\ell; \mathrm{PS}}}{\sqrt{27}M} - 1\,, \nonumber \\
\delta_\gamma &=\ \frac{\gamma_{\mathrm{PS}}}{\gamma_{\ell; \mathrm{Schw}}} - 1 
&&=\ \frac{\gamma_{\mathrm{PS}}}{\pi} - 1\,. 
\end{alignedat}
\end{align}
The first, $\delta$, measures fractional deviations in the shadow diameter of an arbitrary BH from the Schwarzschild value (see, e.g., \cite{EHTC+2022f}). This is not to be confused with the rotation parameter in Refs. \cite{Gralla+2020a, Gralla+2020b}. The shadow diameter, or shadow size, also depends purely on the $tt-$component of the metric alone (see, e.g., Ref. \cite{Psaltis+2020}). The others, $\delta_t$ and $\delta_\gamma$, capture the fractional deviations in the Lyapunov time $t_{\ell; \mathrm{PS}}$ and the lensing Lyapunov exponent $\gamma_{\mathrm{PS}}$ of a BH from the Schwarzschild values. 

\begin{figure*}
\centering
\includegraphics[width=1.9\columnwidth]{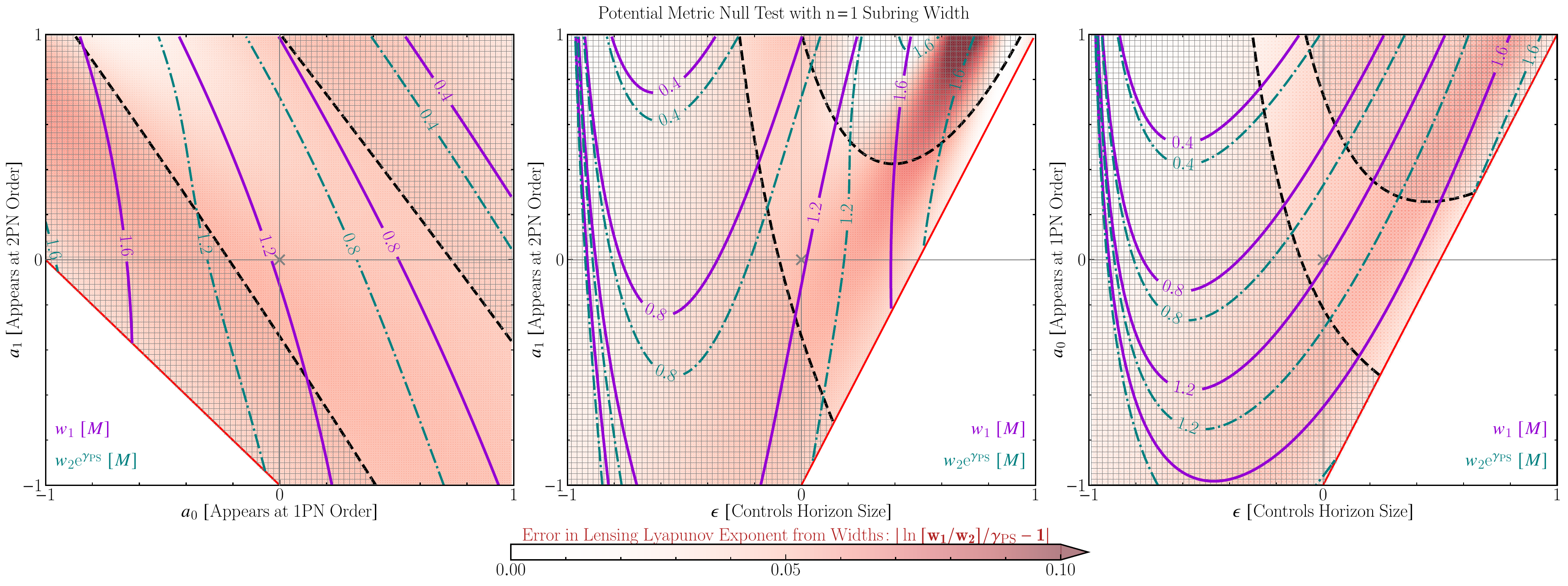}
\caption{\textit{Variation in subring widths with spacetime geometry, for fixed morphology of the emission zone and observer viewing angle.} 
We show the variation in the width of the first subring (solid lines) and the scaled width of the second subring (dot-dashed lines), with changing metric deviation parameters for three different families of Rezzolla-Zhidenko BHs across three different panels. All of these BHs have the same mass $M$, and the Schwarzschild BH is located at $(0, 0)$ in each panel. The hatched regions are disallowed by the 2017 EHT ($1\sigma$) shadow size measurement of Sgr A$^\star$. While it is clear that the shadow size measurement imposes nontrivial constraints on the BH parameter spaces, these extend to infinity in certain directions. Here we demonstrate how, with prior knowledge of the morphology of the emitting region, an additional measurement of the width of the first subring would drastically reduce the allowed band of BH parameters, due to the subring width isocontours being nonparallel to the shadow size isocontours. The morphological parameters for the emission disk here were motivated by state-of-the-art numerical simulations \cite{Narayan+2022}. Finally, the color bars show a typical error of $\lesssim 10\%$ when inferring the lensing Lyapunov exponent using the widths of the first two subrings, across all BH parameter spaces.}
\label{fig:Fig2_RZ_subring_widths}
\end{figure*}

In each panel of Fig. \ref{fig:Fig1_RZ_Pure_Metric_Observables}, we only vary, between $\pm 1$, the metric deviation parameters that are shown on the axes (all others are set to zero). The isocontours corresponding to these different observables do not overlap, in general, and a measurement of any pair of these can constitute a null test of the Schwarzschild metric, in strong gravity, of unprecedented precision. This is especially true when considering the variation in the horizon sizes of nonspinning BHs (see the top-right and bottom-left panels). 

The fractional deviation of the shadow boundary has already been inferred (at the $1\sigma$ level) from the 2017 EHT image of M87$^\star$ to be $\delta = -0.01^{+0.17}_{-0.17}$ \cite{EHTC+2019f, Psaltis+2020} and from the 2017 EHT image of Sgr A$^\star$ to be $\delta = -0.04^{+0.09}_{-0.10}$ (see eq. 12, or Table 2, of Ref. \cite{EHTC+2022c}), albeit with disparate methodologies. These measurements significantly constrain the range of metric deviation parameters, as indicated by the hatched regions for Sgr A$^\star$. However, the allowed bands (non-hatched regions) of the parameter spaces remain noncompact (see also Refs. \cite{Psaltis+2020, Volkel+2021, EHTC+2022f}). For example, it may be possible for $a_0$ to take unboundedly large values,  $a_0 = -\infty$ (for the $(a_0, a_1)-$models) or $a_0 = +\infty$ (for the $(\epsilon, a_0)-$models). This would hold also for the other metric deviation parameters ($\epsilon > -1$ always for BHs however), simply because constraining two-dimensional parameter spaces with a single observable is not, in general, possible. An additional measurement of the lensing Lyapunov exponent can significantly reduce these allowed regions. Indeed, it can render the allowed regions on the (2D) metric deviation parameter spaces compact. 

Furthermore, if such a region does not contain the Schwarzschild values ($\epsilon = a_0 = a_1 = b_0 = b_1 = 0$), then we obtain near-irrefutable evidence of a nonvacuum BH spacetime. This would lead to a violation of one of the assumptions of the Birkhoff theorem, in the strong gravity regime. This could potentially also be interpreted as a precise and accurate smoking-gun signature of a violation of general relativity, as well as of several alternative theories of gravity that admit the Schwarzschild BH metric as a solution. We expect similar statements to become possible even when considering spinning BH spacetimes. Thus, a measurement of the lensing Lyapunov exponent can yield irrefutable evidence of the Kerr metric as being an accurate descriptor of the spacetime geometries of astrophysical ultracompact objects or provide insight into necessary modifications of general relativity in the strong-field regime. 

The BHs considered in the bottom right panel are particularly interesting. As noted above, such BHs have horizons, photon spheres, and ISCOs located at precisely the Schwarzschild BH locations. Furthermore, since they have shadow sizes identical to that of a Schwarzschild BH, deviations in the $rr-$component of the metric due to non-zero $b_0$ or $b_1$ remain completely unconstrained by current EHT measurements. Remember also that the $b_0 = 0$ BHs also pass all solar system constraints, as discussed above. However, as is evident from this panel, both the lensing Lyapunov exponent and the Lyapunov time for these BHs differ from that of a Schwarzschild BH. Thus, inferring either of these critical parameters grants us access to a fundamentally new aspect of the spacetime geometry of astrophysical BHs. 

\section{Constraints on Gravity from Measuring the Width of the First Higher Order Image}
\label{sec:Sec_VI_RZ_Photon_Rings} 

The diameters of photon subrings cast by an emission disk on the observer's sky are tied closely to the shadow diameter of the black hole. Thus, a measurement of a subring diameter yields an excellent additional measurement of the shadow diameter \cite{Gralla+2020b}. Here we consider the impact of a varying spacetime geometry on the \textit{widths} of photon subrings. Since our only purpose here is to demonstrate that widths can play a (surprising) role in metric tests, we will ignore realistic astrophysical effects such as magnetic fields and a varying synchrotron emissivity profile, optical depth, Doppler and gravitational redshifts, etc. Indeed, by ``width'' of the order$-n$ image, here we mean the difference between the lensed image radii of the inner and outer boundaries of the emitting region.  

In particular, we will consider three classes of two-parameter RZ BH metric families, all with $b_0 = b_1 = 0$. We have already seen how current EHT measurements already impose nontrivial constraints on these parameter spaces. We have also discussed how these nonetheless remain unconstrained in certain directions. Here we would like to explore what we can learn additionally about these same parameter spaces from a single additional measurement of the width of a (lowest-order) subring, which is \textit{not} a purely metric-dependent observable.

To cleanly isolate the impact of spacetime geometry, and for simplicity, we fix the observer inclination ($\vartheta_{\mathrm{o}}=0$) and adopt a fiducial configuration for the emission morphology. In Sec. 3.2 of Ref. \cite{Narayan+2022} (see also Refs. \cite{Porth+2019, Chatterjee+2022}), the effective scale-heights $h$ for hot accretion flows around Kerr BHs were studied extensively through general relativistic magentohydrodynamics simulations and it was found that $h/r \lesssim 0.4$. This translates into the faces of the emission zone being located at $(h/2)/r = \tan{[\pm(\pi/2 - \vartheta_{\mathrm{e}})]} \approx \pm[\pi/2 - \vartheta_{\mathrm{e}}]$, or, equivalently, $\vartheta_{\mathrm{e}} \approx \pi/2 \pm 2/10$. Roughly matching their results, we will consider here emission to be sourced from a moderately geometrically-thick disk whose conical faces are at colatitudes of $\vartheta_{\mathrm{e}} = \pi/2 \pm \pi/10$ and whose inner boundary, reasonably, is located at $r_{\mathrm{in}} = r_{\mathrm{H}}$. For concreteness, we pick the outer boundary to be located at $r_{\mathrm{out}} = 3r_{\mathrm{ISCO}}$. Here $r_{\mathrm{ISCO}}$ denotes the location of the timelike Keplerian geodesic that is closest to the BH (see Fig. \ref{fig:FigA1_RZ_Pure_Metric_Characteristics} in the Appendix). Naturally, by increasing (decreasing) the latter, we should expect the widths of all order images to concomitantly increase (decrease). 

In Fig. \ref{fig:Fig2_RZ_subring_widths}, we show the variation in the widths of the first pair of photon subrings with varying metric deviation parameters for the aforementioned fiducial emission region morphological parameters and families of RZ BHs, for a fixed BH mass. We note that the mass of the BH%
\footnote{More accurately, its angular gravitational radius, $\theta_{\mathrm{g}} = (GM/c^2)/D$, where $D$ is the distance to the BH.} %
can be obtained in practice from stellar dynamics measurements (cf., e.g., the discussion in Sec. 9.2 of Ref. \cite{EHTC+2019f} for M87$^\star$ and Sec. 2 of Ref. \cite{EHTC+2022f} for Sgr A$^\star$, and references therein). We uniformly sample the parameter spaces with a resolution of $0.01$ for the $(a_0, a_1)-$models (yielding $\approx\! 35000$ different BHs), $0.01$ for the $(\epsilon, a_1)-$models ($\approx\! 30000$ BHs), and $0.015$ for the $(\epsilon, a_0)-$models ($\approx\! 13000$ BHs). 

The variations in the widths of the $n=1$ subrings across all panels of Fig. \ref{fig:Fig2_RZ_subring_widths} fall roughly within the range $0.3M \lesssim w_1 \lesssim 1.8M$. From the left panel, it appears that the subring widths depend more sensitively on $a_0$ as compared to $a_1$. We understand this to be due to $a_0$ appearing at lower PN order, causing its effect at the photon sphere to be stronger relative to $a_1$. From the remaining panels, it appears that the widths depend more sensitively on $\epsilon$ as compared to either $a_0$ or $a_1$. This is especially true for extremely large BHs ($\epsilon \rightarrow -1$), with larger BHs casting wider subrings. We note that this last trend seems to reverse however (i.e., smaller BHs cast wider subrings) for BHs that have approximately the same event horizon size as a Schwarzschild BH ($\epsilon \approx 0$). These are, of course, only rough trends, and establishing a reliable link between the horizon size and the width of subrings will require a careful disentangling of the possible confounding effects, including the spin of the BH, the choice of parametrization scheme and the various physical effects we have neglected. Nonetheless, these findings uncover an interesting connection between the width of the $n=1$ subring and the spacetime geometry, for a fixed BH mass, and warrant further consideration.

Since the hatched regions in all panels are disallowed by the EHT $1\sigma$ shadow size measurement for Sgr A$^\star$ \cite{EHTC+2022f}, this figure shows then that combining the width of the lowest-order subring with the shadow size measurement, for black holes of known mass, can yield a new and precise null test of the spacetime geometry, due to the orthogonality of their constraints, allowing us to break persisting degeneracies in such BH parameter spaces. 

Finally, we find promising evidence that it may be possible to obtain accurate (with an error $\lesssim 10\%$) inferences of the lensing Lyapunov exponent if a measurement of the widths of a pair of higher-order images ($n=1$ and $n=2$ here) were ever to become possible. As we saw above, this sets up yet another completely new test of the spacetime geometry.

\section{Summary and Discussion}
\label{sec:Sec_VIII_Summary_Discussions}

The recent Event Horizon Telescope images of the supermassive ultracompact objects M87$^\star$ and Sgr A$^\star$ have provided new experimental tests of gravity in the strong-field regime \cite{EHTC+2019f, Psaltis+2020, Kocherlakota+2021, EHTC+2022f, Vagnozzi+2023}. Future imaging observations, including movies, at higher angular resolution and flux sensitivity, are expected to bring the photon ring into focus \cite{Johnson+2020}. Such observations are expected to usher in yet another wave of new experimental tests, as examined here (see also complementary work in Refs. \cite{Gralla+2020b, Wielgus+2020, Ayzenberg2022, Staelens+2023, Broderick+2023, Ayzenberg+2023}).

Central to these tests is the existence of several ``critical parameters,'' that are determined by the spacetime geometry, and which control various properties of the photon ring. The photon ring is located close to the shadow boundary curve (or $n=\infty$ critical curve) on the image plane, and is a collection of higher-order images that are both time-delayed and radially-demagnified versions of the accretion flow. Photons that belong to an image of one higher-order execute approximately an additional half-loop around the BH, causing a time delay between the appearance of different order photons (or images) on the observer's sky. With increasing image order, the time-delay and the radial-demagnification factor become increasingly independent of the properties of the accretion flow \cite{Johnson+2020, Gralla+2020b}. 

Current EHT tests of gravity rely on inferring the size of the shadow (boundary curve). We show here that measuring the characteristic delay time between higher-order images can lead to an independent inference of the same. Recent work has indicated that this may be possible from observations of flaring events associated with Sgr A$^\star$ \cite{Wong2021}. Furthermore, this can also be performed at a multitude of wavelengths, providing a new test of the achromaticity of the shadow, a fundamental prediction of GR. 

The lensing Lyapunov exponent determines the radial-demagnification on the image plane and can be inferred approximately by measuring the diameters or widths of the first pair of photon subrings. Since detecting the $n=2$ subring (tertiary image) may be hard in practice, we find that it can also be inferred from a joint measurement of the delay and Lyapunov times. The latter is a characteristic linear instability timescale for photons present close to the photon sphere to radially-diverge away from it and is also related to the damping timescale of the quasinormal modes of a black hole (cf. Ref. \cite{Cardoso+2009}). It has recently been suggested that this may also be inferred from the late-time behavior of light curves of events involving emitters falling into black holes \cite{Ames+1968, Cardoso+2021}. Finally, a measurement of the width of the $n=1$ subring can also encode nontrivial information about the spacetime geometry. We have shown here how combining measurements of any of the aforementioned observables yields highly nontrivial constraints on black hole parameter spaces. 

By restricting our analysis here to spherically-symmetric and static spacetimes, we were able to obtain the relations between the several spacetime critical parameters rather straightforwardly. However, astrophysical objects are expected to possess nontrivial angular momentum. Therefore, demonstrating extensions of our results to encompass stationary and axisymmetric spacetimes will be hugely exciting for the prospects of experimental gravity with black hole imaging. 

\bigskip

% ############################################################
% Acknowledgements, Bibliography, Appendix
% ############################################################

\begin{acknowledgments}
It is a pleasure to thank Dominic Chang for several insightful discussions and suggestions. We are also grateful to Koushik Chatterjee, Ramesh Narayan, Michael Johnson, Alejandro Cruz-Osorio, and the referee for useful suggestions. PK acknowledges support in part from grants from the Gordon and Betty Moore Foundation and the John Templeton Foundation to the Black Hole Initiative at Harvard University, and from NSF award OISE-1743747. PK and LR acknowledge support from the ERC Advanced Grant ‘JETSET: Launching, propagation and emission of relativistic jets from binary mergers and across mass scales’ (grant no. 884631). LR acknowledges the Walter Greiner Gesellschaft zur F\"orderung der physikalischen Grundlagenforschung e.V. through the Carl W. Fueck Laureatus Chair.
\end{acknowledgments}

\bibliography{Refs-Photon-Rings-Static-BHs-Constraining-Gravity}

\providecommand{\noopsort}[1]{}\providecommand{\singleletter}[1]{#1}%
\begin{thebibliography}{90}%
\makeatletter
\providecommand \@ifxundefined [1]{%
 \@ifx{#1\undefined}
}%
\providecommand \@ifnum [1]{%
 \ifnum #1\expandafter \@firstoftwo
 \else \expandafter \@secondoftwo
 \fi
}%
\providecommand \@ifx [1]{%
 \ifx #1\expandafter \@firstoftwo
 \else \expandafter \@secondoftwo
 \fi
}%
\providecommand \natexlab [1]{#1}%
\providecommand \enquote  [1]{``#1''}%
\providecommand \bibnamefont  [1]{#1}%
\providecommand \bibfnamefont [1]{#1}%
\providecommand \citenamefont [1]{#1}%
\providecommand \href@noop [0]{\@secondoftwo}%
\providecommand \href [0]{\begingroup \@sanitize@url \@href}%
\providecommand \@href[1]{\@@startlink{#1}\@@href}%
\providecommand \@@href[1]{\endgroup#1\@@endlink}%
\providecommand \@sanitize@url [0]{\catcode `\\12\catcode `\$12\catcode
  `\&12\catcode `\#12\catcode `\^12\catcode `\_12\catcode `\%12\relax}%
\providecommand \@@startlink[1]{}%
\providecommand \@@endlink[0]{}%
\providecommand \url  [0]{\begingroup\@sanitize@url \@url }%
\providecommand \@url [1]{\endgroup\@href {#1}{\urlprefix }}%
\providecommand \urlprefix  [0]{URL }%
\providecommand \Eprint [0]{\href }%
\providecommand \doibase [0]{https://doi.org/}%
\providecommand \selectlanguage [0]{\@gobble}%
\providecommand \bibinfo  [0]{\@secondoftwo}%
\providecommand \bibfield  [0]{\@secondoftwo}%
\providecommand \translation [1]{[#1]}%
\providecommand \BibitemOpen [0]{}%
\providecommand \bibitemStop [0]{}%
\providecommand \bibitemNoStop [0]{.\EOS\space}%
\providecommand \EOS [0]{\spacefactor3000\relax}%
\providecommand \BibitemShut  [1]{\csname bibitem#1\endcsname}%
\let\auto@bib@innerbib\@empty
%</preamble>
\bibitem [{\citenamefont {Akiyama}\ \emph
  {et~al.}(2019{\natexlab{a}})\citenamefont {Akiyama} \emph
  {et~al.}}]{EHTC+2019a}%
  \BibitemOpen
  \bibfield  {author} {\bibinfo {author} {\bibfnamefont {K.}~\bibnamefont
  {Akiyama}} \emph {et~al.} (\bibinfo {collaboration} {Event Horizon
  Telescope}),\ }\bibfield  {title} {\bibinfo {title} {{First M87 Event Horizon
  Telescope Results. I. The Shadow of the Supermassive Black Hole}},\ }\href
  {https://doi.org/10.3847/2041-8213/ab0ec7} {\bibfield  {journal} {\bibinfo
  {journal} {Astrophys. J. Lett.}\ }\textbf {\bibinfo {volume} {875}},\
  \bibinfo {pages} {L1} (\bibinfo {year} {2019}{\natexlab{a}})},\ \Eprint
  {https://arxiv.org/abs/1906.11238} {arXiv:1906.11238 [astro-ph.GA]}
  \BibitemShut {NoStop}%
\bibitem [{\citenamefont {Akiyama}\ \emph
  {et~al.}(2022{\natexlab{a}})\citenamefont {Akiyama} \emph
  {et~al.}}]{EHTC+2022a}%
  \BibitemOpen
  \bibfield  {author} {\bibinfo {author} {\bibfnamefont {K.}~\bibnamefont
  {Akiyama}} \emph {et~al.} (\bibinfo {collaboration} {Event Horizon
  Telescope}),\ }\bibfield  {title} {\bibinfo {title} {{First Sagittarius A*
  Event Horizon Telescope Results. I. The Shadow of the Supermassive Black Hole
  in the Center of the Milky Way}},\ }\href
  {https://doi.org/10.3847/2041-8213/ac6674} {\bibfield  {journal} {\bibinfo
  {journal} {Astrophys. J. Lett.}\ }\textbf {\bibinfo {volume} {930}},\
  \bibinfo {pages} {L12} (\bibinfo {year} {2022}{\natexlab{a}})},\ \Eprint
  {https://arxiv.org/abs/2311.08680} {arXiv:2311.08680 [astro-ph.HE]}
  \BibitemShut {NoStop}%
\bibitem [{\citenamefont {{Falcke}}\ \emph {et~al.}(2000)\citenamefont
  {{Falcke}}, \citenamefont {{Melia}},\ and\ \citenamefont
  {{Agol}}}]{Falcke+2000}%
  \BibitemOpen
  \bibfield  {author} {\bibinfo {author} {\bibfnamefont {H.}~\bibnamefont
  {{Falcke}}}, \bibinfo {author} {\bibfnamefont {F.}~\bibnamefont {{Melia}}},\
  and\ \bibinfo {author} {\bibfnamefont {E.}~\bibnamefont {{Agol}}},\
  }\bibfield  {title} {\bibinfo {title} {{Viewing the Shadow of the Black Hole
  at the Galactic Center}},\ }\href {https://doi.org/10.1086/312423} {\bibfield
   {journal} {\bibinfo  {journal} {\apjl}\ }\textbf {\bibinfo {volume} {528}},\
  \bibinfo {pages} {L13} (\bibinfo {year} {2000})},\ \Eprint
  {https://arxiv.org/abs/astro-ph/9912263} {arXiv:astro-ph/9912263 [astro-ph]}
  \BibitemShut {NoStop}%
\bibitem [{\citenamefont {{Yuan}}\ and\ \citenamefont
  {{Narayan}}(2014)}]{Yuan+2014}%
  \BibitemOpen
  \bibfield  {author} {\bibinfo {author} {\bibfnamefont {F.}~\bibnamefont
  {{Yuan}}}\ and\ \bibinfo {author} {\bibfnamefont {R.}~\bibnamefont
  {{Narayan}}},\ }\bibfield  {title} {\bibinfo {title} {{Hot Accretion Flows
  Around Black Holes}},\ }\href
  {https://doi.org/10.1146/annurev-astro-082812-141003} {\bibfield  {journal}
  {\bibinfo  {journal} {\araa}\ }\textbf {\bibinfo {volume} {52}},\ \bibinfo
  {pages} {529} (\bibinfo {year} {2014})},\ \Eprint
  {https://arxiv.org/abs/1401.0586} {arXiv:1401.0586 [astro-ph.HE]}
  \BibitemShut {NoStop}%
\bibitem [{\citenamefont {Akiyama}\ \emph
  {et~al.}(2019{\natexlab{b}})\citenamefont {Akiyama} \emph
  {et~al.}}]{EHTC+2019e}%
  \BibitemOpen
  \bibfield  {author} {\bibinfo {author} {\bibfnamefont {K.}~\bibnamefont
  {Akiyama}} \emph {et~al.} (\bibinfo {collaboration} {Event Horizon
  Telescope}),\ }\bibfield  {title} {\bibinfo {title} {{First M87 Event Horizon
  Telescope Results. V. Physical Origin of the Asymmetric Ring}},\ }\href
  {https://doi.org/10.3847/2041-8213/ab0f43} {\bibfield  {journal} {\bibinfo
  {journal} {Astrophys. J. Lett.}\ }\textbf {\bibinfo {volume} {875}},\
  \bibinfo {pages} {L5} (\bibinfo {year} {2019}{\natexlab{b}})},\ \Eprint
  {https://arxiv.org/abs/1906.11242} {arXiv:1906.11242 [astro-ph.GA]}
  \BibitemShut {NoStop}%
\bibitem [{\citenamefont {Akiyama}\ \emph
  {et~al.}(2022{\natexlab{b}})\citenamefont {Akiyama} \emph
  {et~al.}}]{EHTC+2022d}%
  \BibitemOpen
  \bibfield  {author} {\bibinfo {author} {\bibfnamefont {K.}~\bibnamefont
  {Akiyama}} \emph {et~al.} (\bibinfo {collaboration} {Event Horizon
  Telescope}),\ }\bibfield  {title} {\bibinfo {title} {{First Sagittarius A*
  Event Horizon Telescope Results. IV. Variability, Morphology, and Black Hole
  Mass}},\ }\href {https://doi.org/10.3847/2041-8213/ac6736} {\bibfield
  {journal} {\bibinfo  {journal} {Astrophys. J. Lett.}\ }\textbf {\bibinfo
  {volume} {930}},\ \bibinfo {pages} {L15} (\bibinfo {year}
  {2022}{\natexlab{b}})},\ \Eprint {https://arxiv.org/abs/2311.08697}
  {arXiv:2311.08697 [astro-ph.HE]} \BibitemShut {NoStop}%
\bibitem [{\citenamefont {Akiyama}\ \emph
  {et~al.}(2022{\natexlab{c}})\citenamefont {Akiyama} \emph
  {et~al.}}]{EHTC+2022f}%
  \BibitemOpen
  \bibfield  {author} {\bibinfo {author} {\bibfnamefont {K.}~\bibnamefont
  {Akiyama}} \emph {et~al.} (\bibinfo {collaboration} {Event Horizon
  Telescope}),\ }\bibfield  {title} {\bibinfo {title} {{First Sagittarius A*
  Event Horizon Telescope Results. VI. Testing the Black Hole Metric}},\ }\href
  {https://doi.org/10.3847/2041-8213/ac6756} {\bibfield  {journal} {\bibinfo
  {journal} {Astrophys. J. Lett.}\ }\textbf {\bibinfo {volume} {930}},\
  \bibinfo {pages} {L17} (\bibinfo {year} {2022}{\natexlab{c}})},\ \Eprint
  {https://arxiv.org/abs/2311.09484} {arXiv:2311.09484 [astro-ph.HE]}
  \BibitemShut {NoStop}%
\bibitem [{\citenamefont {{Cunha}}\ and\ \citenamefont
  {{Herdeiro}}(2020)}]{Cunha+2020}%
  \BibitemOpen
  \bibfield  {author} {\bibinfo {author} {\bibfnamefont {P.~V.~P.}\
  \bibnamefont {{Cunha}}}\ and\ \bibinfo {author} {\bibfnamefont {C.~A.~R.}\
  \bibnamefont {{Herdeiro}}},\ }\bibfield  {title} {\bibinfo {title}
  {{Stationary Black Holes and Light Rings}},\ }\href
  {https://doi.org/10.1103/PhysRevLett.124.181101} {\bibfield  {journal}
  {\bibinfo  {journal} {\prl}\ }\textbf {\bibinfo {volume} {124}},\ \bibinfo
  {eid} {181101} (\bibinfo {year} {2020})},\ \Eprint
  {https://arxiv.org/abs/2003.06445} {arXiv:2003.06445 [gr-qc]} \BibitemShut
  {NoStop}%
\bibitem [{\citenamefont {{Ghosh}}\ and\ \citenamefont
  {{Sarkar}}(2021)}]{Ghosh+2021}%
  \BibitemOpen
  \bibfield  {author} {\bibinfo {author} {\bibfnamefont {R.}~\bibnamefont
  {{Ghosh}}}\ and\ \bibinfo {author} {\bibfnamefont {S.}~\bibnamefont
  {{Sarkar}}},\ }\bibfield  {title} {\bibinfo {title} {{Light rings of
  stationary spacetimes}},\ }\href
  {https://doi.org/10.1103/PhysRevD.104.044019} {\bibfield  {journal} {\bibinfo
   {journal} {\prd}\ }\textbf {\bibinfo {volume} {104}},\ \bibinfo {eid}
  {044019} (\bibinfo {year} {2021})},\ \Eprint
  {https://arxiv.org/abs/2107.07370} {arXiv:2107.07370 [gr-qc]} \BibitemShut
  {NoStop}%
\bibitem [{\citenamefont {{Bardeen}}(1973)}]{Bardeen1973}%
  \BibitemOpen
  \bibfield  {author} {\bibinfo {author} {\bibfnamefont {J.~M.}\ \bibnamefont
  {{Bardeen}}},\ }\bibfield  {title} {\bibinfo {title} {{Timelike and null
  geodesics in the Kerr metric.}},\ }in\ \href
  {https://adsabs.harvard.edu/full/1974IAUS...64..132B} {\emph {\bibinfo
  {booktitle} {Black Holes (Les Astres Occlus)}}}\ (\bibinfo {year} {1973})\
  pp.\ \bibinfo {pages} {215--239}\BibitemShut {NoStop}%
\bibitem [{\citenamefont {{Johannsen}}\ and\ \citenamefont
  {{Psaltis}}(2010)}]{Johannsen+2010}%
  \BibitemOpen
  \bibfield  {author} {\bibinfo {author} {\bibfnamefont {T.}~\bibnamefont
  {{Johannsen}}}\ and\ \bibinfo {author} {\bibfnamefont {D.}~\bibnamefont
  {{Psaltis}}},\ }\bibfield  {title} {\bibinfo {title} {{Testing the No-hair
  Theorem with Observations in the Electromagnetic Spectrum. II. Black Hole
  Images}},\ }\href {https://doi.org/10.1088/0004-637X/718/1/446} {\bibfield
  {journal} {\bibinfo  {journal} {\apj}\ }\textbf {\bibinfo {volume} {718}},\
  \bibinfo {pages} {446} (\bibinfo {year} {2010})},\ \Eprint
  {https://arxiv.org/abs/1005.1931} {arXiv:1005.1931 [astro-ph.HE]}
  \BibitemShut {NoStop}%
\bibitem [{\citenamefont {Akiyama}\ \emph
  {et~al.}(2019{\natexlab{c}})\citenamefont {Akiyama} \emph
  {et~al.}}]{EHTC+2019f}%
  \BibitemOpen
  \bibfield  {author} {\bibinfo {author} {\bibfnamefont {K.}~\bibnamefont
  {Akiyama}} \emph {et~al.} (\bibinfo {collaboration} {Event Horizon
  Telescope}),\ }\bibfield  {title} {\bibinfo {title} {{First M87 Event Horizon
  Telescope Results. VI. The Shadow and Mass of the Central Black Hole}},\
  }\href {https://doi.org/10.3847/2041-8213/ab1141} {\bibfield  {journal}
  {\bibinfo  {journal} {Astrophys. J. Lett.}\ }\textbf {\bibinfo {volume}
  {875}},\ \bibinfo {pages} {L6} (\bibinfo {year} {2019}{\natexlab{c}})},\
  \Eprint {https://arxiv.org/abs/1906.11243} {arXiv:1906.11243 [astro-ph.GA]}
  \BibitemShut {NoStop}%
\bibitem [{\citenamefont {Psaltis}\ \emph {et~al.}(2020)\citenamefont {Psaltis}
  \emph {et~al.}}]{Psaltis+2020}%
  \BibitemOpen
  \bibfield  {author} {\bibinfo {author} {\bibfnamefont {D.}~\bibnamefont
  {Psaltis}} \emph {et~al.} (\bibinfo {collaboration} {Event Horizon
  Telescope}),\ }\bibfield  {title} {\bibinfo {title} {{Gravitational Test
  Beyond the First Post-Newtonian Order with the Shadow of the M87 Black
  Hole}},\ }\href {https://doi.org/10.1103/PhysRevLett.125.141104} {\bibfield
  {journal} {\bibinfo  {journal} {Phys. Rev. Lett.}\ }\textbf {\bibinfo
  {volume} {125}},\ \bibinfo {pages} {141104} (\bibinfo {year} {2020})},\
  \Eprint {https://arxiv.org/abs/2010.01055} {arXiv:2010.01055 [gr-qc]}
  \BibitemShut {NoStop}%
\bibitem [{\citenamefont {Kocherlakota}\ \emph {et~al.}(2021)\citenamefont
  {Kocherlakota} \emph {et~al.}}]{Kocherlakota+2021}%
  \BibitemOpen
  \bibfield  {author} {\bibinfo {author} {\bibfnamefont {P.}~\bibnamefont
  {Kocherlakota}} \emph {et~al.} (\bibinfo {collaboration} {Event Horizon
  Telescope}),\ }\bibfield  {title} {\bibinfo {title} {{Constraints on
  black-hole charges with the 2017 EHT observations of M87*}},\ }\href
  {https://doi.org/10.1103/PhysRevD.103.104047} {\bibfield  {journal} {\bibinfo
   {journal} {Phys. Rev. D}\ }\textbf {\bibinfo {volume} {103}},\ \bibinfo
  {pages} {104047} (\bibinfo {year} {2021})},\ \Eprint
  {https://arxiv.org/abs/2105.09343} {arXiv:2105.09343 [gr-qc]} \BibitemShut
  {NoStop}%
\bibitem [{\citenamefont {Abbott}\ \emph {et~al.}(2019)\citenamefont {Abbott}
  \emph {et~al.}}]{LSC+2019}%
  \BibitemOpen
  \bibfield  {author} {\bibinfo {author} {\bibfnamefont {B.~P.}\ \bibnamefont
  {Abbott}} \emph {et~al.} (\bibinfo {collaboration} {LIGO Scientific,
  Virgo}),\ }\bibfield  {title} {\bibinfo {title} {{Tests of General Relativity
  with the Binary Black Hole Signals from the LIGO-Virgo Catalog GWTC-1}},\
  }\href {https://doi.org/10.1103/PhysRevD.100.104036} {\bibfield  {journal}
  {\bibinfo  {journal} {Phys. Rev. D}\ }\textbf {\bibinfo {volume} {100}},\
  \bibinfo {pages} {104036} (\bibinfo {year} {2019})},\ \Eprint
  {https://arxiv.org/abs/1903.04467} {arXiv:1903.04467 [gr-qc]} \BibitemShut
  {NoStop}%
\bibitem [{\citenamefont {Abbott}\ \emph {et~al.}(2021)\citenamefont {Abbott}
  \emph {et~al.}}]{LSC+2021a}%
  \BibitemOpen
  \bibfield  {author} {\bibinfo {author} {\bibfnamefont {R.}~\bibnamefont
  {Abbott}} \emph {et~al.} (\bibinfo {collaboration} {LIGO Scientific,
  Virgo}),\ }\bibfield  {title} {\bibinfo {title} {{Tests of general relativity
  with binary black holes from the second LIGO-Virgo gravitational-wave
  transient catalog}},\ }\href {https://doi.org/10.1103/PhysRevD.103.122002}
  {\bibfield  {journal} {\bibinfo  {journal} {Phys. Rev. D}\ }\textbf {\bibinfo
  {volume} {103}},\ \bibinfo {pages} {122002} (\bibinfo {year} {2021})},\
  \Eprint {https://arxiv.org/abs/2010.14529} {arXiv:2010.14529 [gr-qc]}
  \BibitemShut {NoStop}%
\bibitem [{\citenamefont {Abbott}\ \emph {et~al.}()\citenamefont {Abbott} \emph
  {et~al.}}]{LSC+2021b}%
  \BibitemOpen
  \bibfield  {author} {\bibinfo {author} {\bibfnamefont {R.}~\bibnamefont
  {Abbott}} \emph {et~al.} (\bibinfo {collaboration} {LIGO Scientific, VIRGO,
  KAGRA}),\ }\bibfield  {title} {\bibinfo {title} {{Tests of General Relativity
  with GWTC-3}},\ }\href@noop {} {\ }\Eprint {https://arxiv.org/abs/2112.06861}
  {arXiv:2112.06861 [gr-qc]} \BibitemShut {NoStop}%
\bibitem [{\citenamefont {{V{\"o}lkel}}\ \emph {et~al.}(2021)\citenamefont
  {{V{\"o}lkel}}, \citenamefont {{Barausse}}, \citenamefont {{Franchini}},\
  and\ \citenamefont {{Broderick}}}]{Volkel+2021}%
  \BibitemOpen
  \bibfield  {author} {\bibinfo {author} {\bibfnamefont {S.~H.}\ \bibnamefont
  {{V{\"o}lkel}}}, \bibinfo {author} {\bibfnamefont {E.}~\bibnamefont
  {{Barausse}}}, \bibinfo {author} {\bibfnamefont {N.}~\bibnamefont
  {{Franchini}}},\ and\ \bibinfo {author} {\bibfnamefont {A.~E.}\ \bibnamefont
  {{Broderick}}},\ }\bibfield  {title} {\bibinfo {title} {{EHT tests of the
  strong-field regime of general relativity}},\ }\href
  {https://doi.org/10.1088/1361-6382/ac27ed} {\bibfield  {journal} {\bibinfo
  {journal} {\cqg}\ }\textbf {\bibinfo {volume} {38}},\ \bibinfo {eid} {21LT01}
  (\bibinfo {year} {2021})},\ \Eprint {https://arxiv.org/abs/2011.06812}
  {arXiv:2011.06812 [gr-qc]} \BibitemShut {NoStop}%
\bibitem [{\citenamefont {{Kocherlakota}}\ and\ \citenamefont
  {{Rezzolla}}(2022)}]{Kocherlakota+2022}%
  \BibitemOpen
  \bibfield  {author} {\bibinfo {author} {\bibfnamefont {P.}~\bibnamefont
  {{Kocherlakota}}}\ and\ \bibinfo {author} {\bibfnamefont {L.}~\bibnamefont
  {{Rezzolla}}},\ }\bibfield  {title} {\bibinfo {title} {{Distinguishing
  gravitational and emission physics in black hole imaging: spherical
  symmetry}},\ }\href {https://doi.org/10.1093/mnras/stac891} {\bibfield
  {journal} {\bibinfo  {journal} {\mnras}\ }\textbf {\bibinfo {volume} {513}},\
  \bibinfo {pages} {1229} (\bibinfo {year} {2022})},\ \Eprint
  {https://arxiv.org/abs/2201.05641} {arXiv:2201.05641 [gr-qc]} \BibitemShut
  {NoStop}%
\bibitem [{\citenamefont {Kumar}\ \emph {et~al.}(2020)\citenamefont {Kumar},
  \citenamefont {Kumar},\ and\ \citenamefont {Ghosh}}]{Kumar+2020}%
  \BibitemOpen
  \bibfield  {author} {\bibinfo {author} {\bibfnamefont {R.}~\bibnamefont
  {Kumar}}, \bibinfo {author} {\bibfnamefont {A.}~\bibnamefont {Kumar}},\ and\
  \bibinfo {author} {\bibfnamefont {S.~G.}\ \bibnamefont {Ghosh}},\ }\bibfield
  {title} {\bibinfo {title} {{Testing Rotating Regular Metrics as Candidates
  for Astrophysical Black Holes}},\ }\href
  {https://doi.org/10.3847/1538-4357/ab8c4a} {\bibfield  {journal} {\bibinfo
  {journal} {Astrophys. J.}\ }\textbf {\bibinfo {volume} {896}},\ \bibinfo
  {pages} {89} (\bibinfo {year} {2020})},\ \Eprint
  {https://arxiv.org/abs/2006.09869} {arXiv:2006.09869 [gr-qc]} \BibitemShut
  {NoStop}%
\bibitem [{\citenamefont {Ghosh}\ and\ \citenamefont
  {walia}(2021)}]{Ghosh+2021b}%
  \BibitemOpen
  \bibfield  {author} {\bibinfo {author} {\bibfnamefont {S.~G.}\ \bibnamefont
  {Ghosh}}\ and\ \bibinfo {author} {\bibfnamefont {R.~K.}\ \bibnamefont
  {walia}},\ }\bibfield  {title} {\bibinfo {title} {{Rotating black holes in
  general relativity coupled to nonlinear electrodynamics}},\ }\href
  {https://doi.org/10.1016/j.aop.2021.168619} {\bibfield  {journal} {\bibinfo
  {journal} {Annals Phys.}\ }\textbf {\bibinfo {volume} {434}},\ \bibinfo
  {pages} {168619} (\bibinfo {year} {2021})},\ \Eprint
  {https://arxiv.org/abs/2109.13031} {arXiv:2109.13031 [gr-qc]} \BibitemShut
  {NoStop}%
\bibitem [{\citenamefont {{Vagnozzi}}\ \emph {et~al.}(2023)\citenamefont
  {{Vagnozzi}} \emph {et~al.}}]{Vagnozzi+2023}%
  \BibitemOpen
  \bibfield  {author} {\bibinfo {author} {\bibfnamefont {S.}~\bibnamefont
  {{Vagnozzi}}} \emph {et~al.},\ }\bibfield  {title} {\bibinfo {title}
  {{Horizon-scale tests of gravity theories and fundamental physics from the
  Event Horizon Telescope image of Sagittarius A*}},\ }\href
  {https://doi.org/10.1088/1361-6382/acd97b} {\bibfield  {journal} {\bibinfo
  {journal} {\cqg}\ }\textbf {\bibinfo {volume} {40}},\ \bibinfo {eid} {165007}
  (\bibinfo {year} {2023})},\ \Eprint {https://arxiv.org/abs/2205.07787}
  {arXiv:2205.07787 [gr-qc]} \BibitemShut {NoStop}%
\bibitem [{\citenamefont {{Teo}}(1998)}]{Teo1998}%
  \BibitemOpen
  \bibfield  {author} {\bibinfo {author} {\bibfnamefont {E.}~\bibnamefont
  {{Teo}}},\ }\bibfield  {title} {\bibinfo {title} {{Rotating traversable
  wormholes}},\ }\href {https://doi.org/10.1103/PhysRevD.58.024014} {\bibfield
  {journal} {\bibinfo  {journal} {\prd}\ }\textbf {\bibinfo {volume} {58}},\
  \bibinfo {eid} {024014} (\bibinfo {year} {1998})},\ \Eprint
  {https://arxiv.org/abs/gr-qc/9803098} {arXiv:gr-qc/9803098 [gr-qc]}
  \BibitemShut {NoStop}%
\bibitem [{\citenamefont {{Hioki}}\ and\ \citenamefont
  {{Miyamoto}}(2008)}]{Hioki2008}%
  \BibitemOpen
  \bibfield  {author} {\bibinfo {author} {\bibfnamefont {K.}~\bibnamefont
  {{Hioki}}}\ and\ \bibinfo {author} {\bibfnamefont {U.}~\bibnamefont
  {{Miyamoto}}},\ }\bibfield  {title} {\bibinfo {title} {{Hidden symmetries,
  null geodesics, and photon capture in the Sen black hole}},\ }\href
  {https://doi.org/10.1103/PhysRevD.78.044007} {\bibfield  {journal} {\bibinfo
  {journal} {\prd}\ }\textbf {\bibinfo {volume} {78}},\ \bibinfo {eid} {044007}
  (\bibinfo {year} {2008})},\ \Eprint {https://arxiv.org/abs/0805.3146}
  {arXiv:0805.3146 [gr-qc]} \BibitemShut {NoStop}%
\bibitem [{\citenamefont {{Chowdhury}}\ \emph {et~al.}(2012)\citenamefont
  {{Chowdhury}}, \citenamefont {{Patil}}, \citenamefont {{Malafarina}},\ and\
  \citenamefont {{Joshi}}}]{Chowdhury+2012}%
  \BibitemOpen
  \bibfield  {author} {\bibinfo {author} {\bibfnamefont {A.~N.}\ \bibnamefont
  {{Chowdhury}}}, \bibinfo {author} {\bibfnamefont {M.}~\bibnamefont
  {{Patil}}}, \bibinfo {author} {\bibfnamefont {D.}~\bibnamefont
  {{Malafarina}}},\ and\ \bibinfo {author} {\bibfnamefont {P.~S.}\ \bibnamefont
  {{Joshi}}},\ }\bibfield  {title} {\bibinfo {title} {{Circular geodesics and
  accretion disks in the Janis-Newman-Winicour and gamma metric spacetimes}},\
  }\href {https://doi.org/10.1103/PhysRevD.85.104031} {\bibfield  {journal}
  {\bibinfo  {journal} {\prd}\ }\textbf {\bibinfo {volume} {85}},\ \bibinfo
  {eid} {104031} (\bibinfo {year} {2012})},\ \Eprint
  {https://arxiv.org/abs/1112.2522} {arXiv:1112.2522 [gr-qc]} \BibitemShut
  {NoStop}%
\bibitem [{\citenamefont {{Sakai}}\ \emph {et~al.}(2014)\citenamefont
  {{Sakai}}, \citenamefont {{Saida}},\ and\ \citenamefont
  {{Tamaki}}}]{Sakai+2014}%
  \BibitemOpen
  \bibfield  {author} {\bibinfo {author} {\bibfnamefont {N.}~\bibnamefont
  {{Sakai}}}, \bibinfo {author} {\bibfnamefont {H.}~\bibnamefont {{Saida}}},\
  and\ \bibinfo {author} {\bibfnamefont {T.}~\bibnamefont {{Tamaki}}},\
  }\bibfield  {title} {\bibinfo {title} {{Gravastar shadows}},\ }\href
  {https://doi.org/10.1103/PhysRevD.90.104013} {\bibfield  {journal} {\bibinfo
  {journal} {\prd}\ }\textbf {\bibinfo {volume} {90}},\ \bibinfo {eid} {104013}
  (\bibinfo {year} {2014})},\ \Eprint {https://arxiv.org/abs/1408.6929}
  {arXiv:1408.6929 [gr-qc]} \BibitemShut {NoStop}%
\bibitem [{\citenamefont {Cunha}\ \emph {et~al.}(2016)\citenamefont {Cunha}
  \emph {et~al.}}]{Cunha+2016}%
  \BibitemOpen
  \bibfield  {author} {\bibinfo {author} {\bibfnamefont {P.}~\bibnamefont
  {Cunha}} \emph {et~al.},\ }\bibfield  {title} {\bibinfo {title} {{Chaotic
  lensing around boson stars and Kerr black holes with scalar hair}},\ }\href
  {https://doi.org/10.1103/PhysRevD.94.104023} {\bibfield  {journal} {\bibinfo
  {journal} {Phys. Rev. D}\ }\textbf {\bibinfo {volume} {94}},\ \bibinfo
  {pages} {104023} (\bibinfo {year} {2016})},\ \Eprint
  {https://arxiv.org/abs/1609.01340} {arXiv:1609.01340 [gr-qc]} \BibitemShut
  {NoStop}%
\bibitem [{\citenamefont {{Grandcl{\'e}ment}}(2017)}]{Grandclement2017}%
  \BibitemOpen
  \bibfield  {author} {\bibinfo {author} {\bibfnamefont {P.}~\bibnamefont
  {{Grandcl{\'e}ment}}},\ }\bibfield  {title} {\bibinfo {title} {{Light rings
  and light points of boson stars}},\ }\href
  {https://doi.org/10.1103/PhysRevD.95.084011} {\bibfield  {journal} {\bibinfo
  {journal} {\prd}\ }\textbf {\bibinfo {volume} {95}},\ \bibinfo {eid} {084011}
  (\bibinfo {year} {2017})},\ \Eprint {https://arxiv.org/abs/1612.07507}
  {arXiv:1612.07507 [gr-qc]} \BibitemShut {NoStop}%
\bibitem [{\citenamefont {Shaikh}(2018)}]{Shaikh2018}%
  \BibitemOpen
  \bibfield  {author} {\bibinfo {author} {\bibfnamefont {R.}~\bibnamefont
  {Shaikh}},\ }\bibfield  {title} {\bibinfo {title} {{Shadows of rotating
  wormholes}},\ }\href {https://doi.org/10.1103/PhysRevD.98.024044} {\bibfield
  {journal} {\bibinfo  {journal} {Phys. Rev. D}\ }\textbf {\bibinfo {volume}
  {98}},\ \bibinfo {pages} {024044} (\bibinfo {year} {2018})},\ \Eprint
  {https://arxiv.org/abs/1803.11422} {arXiv:1803.11422 [gr-qc]} \BibitemShut
  {NoStop}%
\bibitem [{\citenamefont {{Shaikh}}\ \emph {et~al.}(2019)\citenamefont
  {{Shaikh}}, \citenamefont {{Kocherlakota}}, \citenamefont {{Narayan}},\ and\
  \citenamefont {{Joshi}}}]{Shaikh+2019}%
  \BibitemOpen
  \bibfield  {author} {\bibinfo {author} {\bibfnamefont {R.}~\bibnamefont
  {{Shaikh}}}, \bibinfo {author} {\bibfnamefont {P.}~\bibnamefont
  {{Kocherlakota}}}, \bibinfo {author} {\bibfnamefont {R.}~\bibnamefont
  {{Narayan}}},\ and\ \bibinfo {author} {\bibfnamefont {P.~S.}\ \bibnamefont
  {{Joshi}}},\ }\bibfield  {title} {\bibinfo {title} {{Shadows of spherically
  symmetric black holes and naked singularities}},\ }\href
  {https://doi.org/10.1093/mnras/sty2624} {\bibfield  {journal} {\bibinfo
  {journal} {\mnras}\ }\textbf {\bibinfo {volume} {482}},\ \bibinfo {pages}
  {52} (\bibinfo {year} {2019})},\ \Eprint {https://arxiv.org/abs/1802.08060}
  {arXiv:1802.08060 [astro-ph.HE]} \BibitemShut {NoStop}%
\bibitem [{\citenamefont {{Rahaman}}\ \emph {et~al.}(2021)\citenamefont
  {{Rahaman}}, \citenamefont {{Manna}}, \citenamefont {{Shaikh}}, \citenamefont
  {{Aktar}}, \citenamefont {{Mondal}},\ and\ \citenamefont
  {{Samanta}}}]{Rahaman+2021}%
  \BibitemOpen
  \bibfield  {author} {\bibinfo {author} {\bibfnamefont {F.}~\bibnamefont
  {{Rahaman}}}, \bibinfo {author} {\bibfnamefont {T.}~\bibnamefont {{Manna}}},
  \bibinfo {author} {\bibfnamefont {R.}~\bibnamefont {{Shaikh}}}, \bibinfo
  {author} {\bibfnamefont {S.}~\bibnamefont {{Aktar}}}, \bibinfo {author}
  {\bibfnamefont {M.}~\bibnamefont {{Mondal}}},\ and\ \bibinfo {author}
  {\bibfnamefont {B.}~\bibnamefont {{Samanta}}},\ }\bibfield  {title} {\bibinfo
  {title} {{Thin accretion disks around traversable wormholes}},\ }\href
  {https://doi.org/10.48550/arXiv.2110.09820} {\bibfield  {journal} {\bibinfo
  {journal} {arXiv e-prints}\ ,\ \bibinfo {eid} {arXiv:2110.09820}} (\bibinfo
  {year} {2021})},\ \Eprint {https://arxiv.org/abs/2110.09820}
  {arXiv:2110.09820 [gr-qc]} \BibitemShut {NoStop}%
\bibitem [{\citenamefont {Shaikh}\ \emph {et~al.}(2021)\citenamefont {Shaikh},
  \citenamefont {Pal}, \citenamefont {Pal},\ and\ \citenamefont
  {Sarkar}}]{Shaikh+2021}%
  \BibitemOpen
  \bibfield  {author} {\bibinfo {author} {\bibfnamefont {R.}~\bibnamefont
  {Shaikh}}, \bibinfo {author} {\bibfnamefont {K.}~\bibnamefont {Pal}},
  \bibinfo {author} {\bibfnamefont {K.}~\bibnamefont {Pal}},\ and\ \bibinfo
  {author} {\bibfnamefont {T.}~\bibnamefont {Sarkar}},\ }\bibfield  {title}
  {\bibinfo {title} {{Constraining alternatives to the Kerr black hole}},\
  }\href {https://doi.org/10.1093/mnras/stab1779} {\bibfield  {journal}
  {\bibinfo  {journal} {Mon. Not. Roy. Astron. Soc.}\ }\textbf {\bibinfo
  {volume} {506}},\ \bibinfo {pages} {1229} (\bibinfo {year} {2021})},\ \Eprint
  {https://arxiv.org/abs/2102.04299} {arXiv:2102.04299 [gr-qc]} \BibitemShut
  {NoStop}%
\bibitem [{\citenamefont {{Solanki}}\ \emph {et~al.}(2022)\citenamefont
  {{Solanki}} \emph {et~al.}}]{Solanki+2022}%
  \BibitemOpen
  \bibfield  {author} {\bibinfo {author} {\bibfnamefont {D.~N.}\ \bibnamefont
  {{Solanki}}} \emph {et~al.},\ }\bibfield  {title} {\bibinfo {title} {{Shadows
  and precession of orbits in rotating Janis-Newman-Winicour spacetime}},\
  }\href {https://doi.org/10.1140/epjc/s10052-022-10045-1} {\bibfield
  {journal} {\bibinfo  {journal} {\epjc}\ }\textbf {\bibinfo {volume} {82}},\
  \bibinfo {eid} {77} (\bibinfo {year} {2022})},\ \Eprint
  {https://arxiv.org/abs/2109.14937} {arXiv:2109.14937 [gr-qc]} \BibitemShut
  {NoStop}%
\bibitem [{\citenamefont {Wald}(1997)}]{Wald1997}%
  \BibitemOpen
  \bibfield  {author} {\bibinfo {author} {\bibfnamefont {R.~M.}\ \bibnamefont
  {Wald}},\ }\bibfield  {title} {\bibinfo {title} {{Gravitational collapse and
  cosmic censorship}}\ }(\bibinfo {year} {1997})\ pp.\ \bibinfo {pages}
  {69--85},\ \Eprint {https://arxiv.org/abs/gr-qc/9710068}
  {arXiv:gr-qc/9710068} \BibitemShut {NoStop}%
\bibitem [{\citenamefont {Kumar}\ and\ \citenamefont
  {Ghosh}(2020)}]{Kumar+2020b}%
  \BibitemOpen
  \bibfield  {author} {\bibinfo {author} {\bibfnamefont {R.}~\bibnamefont
  {Kumar}}\ and\ \bibinfo {author} {\bibfnamefont {S.~G.}\ \bibnamefont
  {Ghosh}},\ }\bibfield  {title} {\bibinfo {title} {{Rotating black holes in
  $4D$ Einstein-Gauss-Bonnet gravity and its shadow}},\ }\href
  {https://doi.org/10.1088/1475-7516/2020/07/053} {\bibfield  {journal}
  {\bibinfo  {journal} {JCAP}\ }\textbf {\bibinfo {volume} {07}},\ \bibinfo
  {pages} {053}},\ \Eprint {https://arxiv.org/abs/2003.08927} {arXiv:2003.08927
  [gr-qc]} \BibitemShut {NoStop}%
\bibitem [{\citenamefont {Afrin}\ \emph {et~al.}(2021)\citenamefont {Afrin},
  \citenamefont {Kumar},\ and\ \citenamefont {Ghosh}}]{Afrin+2021}%
  \BibitemOpen
  \bibfield  {author} {\bibinfo {author} {\bibfnamefont {M.}~\bibnamefont
  {Afrin}}, \bibinfo {author} {\bibfnamefont {R.}~\bibnamefont {Kumar}},\ and\
  \bibinfo {author} {\bibfnamefont {S.~G.}\ \bibnamefont {Ghosh}},\ }\bibfield
  {title} {\bibinfo {title} {{Parameter estimation of hairy Kerr black holes
  from its shadow and constraints from M87*}},\ }\href
  {https://doi.org/10.1093/mnras/stab1260} {\bibfield  {journal} {\bibinfo
  {journal} {Mon. Not. Roy. Astron. Soc.}\ }\textbf {\bibinfo {volume} {504}},\
  \bibinfo {pages} {5927} (\bibinfo {year} {2021})},\ \Eprint
  {https://arxiv.org/abs/2103.11417} {arXiv:2103.11417 [gr-qc]} \BibitemShut
  {NoStop}%
\bibitem [{\citenamefont {Kumar~Walia}(2023)}]{Kumar+2022}%
  \BibitemOpen
  \bibfield  {author} {\bibinfo {author} {\bibfnamefont {R.}~\bibnamefont
  {Kumar~Walia}},\ }\bibfield  {title} {\bibinfo {title} {{Observational
  predictions of LQG motivated polymerized black holes and constraints from Sgr
  A* and M87*}},\ }\href {https://doi.org/10.1088/1475-7516/2023/03/029}
  {\bibfield  {journal} {\bibinfo  {journal} {JCAP}\ }\textbf {\bibinfo
  {volume} {03}},\ \bibinfo {pages} {029}},\ \Eprint
  {https://arxiv.org/abs/2207.02106} {arXiv:2207.02106 [gr-qc]} \BibitemShut
  {NoStop}%
\bibitem [{\citenamefont {{Gralla}}\ \emph {et~al.}(2020)\citenamefont
  {{Gralla}}, \citenamefont {{Lupsasca}},\ and\ \citenamefont
  {{Marrone}}}]{Gralla+2020b}%
  \BibitemOpen
  \bibfield  {author} {\bibinfo {author} {\bibfnamefont {S.~E.}\ \bibnamefont
  {{Gralla}}}, \bibinfo {author} {\bibfnamefont {A.}~\bibnamefont
  {{Lupsasca}}},\ and\ \bibinfo {author} {\bibfnamefont {D.~P.}\ \bibnamefont
  {{Marrone}}},\ }\bibfield  {title} {\bibinfo {title} {{The shape of the black
  hole photon ring: A precise test of strong-field general relativity}},\
  }\href {https://doi.org/10.1103/PhysRevD.102.124004} {\bibfield  {journal}
  {\bibinfo  {journal} {\prd}\ }\textbf {\bibinfo {volume} {102}},\ \bibinfo
  {eid} {124004} (\bibinfo {year} {2020})},\ \Eprint
  {https://arxiv.org/abs/2008.03879} {arXiv:2008.03879 [gr-qc]} \BibitemShut
  {NoStop}%
\bibitem [{\citenamefont {{Wielgus}}(2021)}]{Wielgus2021}%
  \BibitemOpen
  \bibfield  {author} {\bibinfo {author} {\bibfnamefont {M.}~\bibnamefont
  {{Wielgus}}},\ }\bibfield  {title} {\bibinfo {title} {{Photon rings of
  spherically symmetric black holes and robust tests of non-Kerr metrics}},\
  }\href {https://doi.org/10.1103/PhysRevD.104.124058} {\bibfield  {journal}
  {\bibinfo  {journal} {\prd}\ }\textbf {\bibinfo {volume} {104}},\ \bibinfo
  {eid} {124058} (\bibinfo {year} {2021})},\ \Eprint
  {https://arxiv.org/abs/2109.10840} {arXiv:2109.10840 [gr-qc]} \BibitemShut
  {NoStop}%
\bibitem [{\citenamefont {Ayzenberg}(2022)}]{Ayzenberg2022}%
  \BibitemOpen
  \bibfield  {author} {\bibinfo {author} {\bibfnamefont {D.}~\bibnamefont
  {Ayzenberg}},\ }\bibfield  {title} {\bibinfo {title} {{Testing gravity with
  black hole shadow subrings}},\ }\href
  {https://doi.org/10.1088/1361-6382/ac655d} {\bibfield  {journal} {\bibinfo
  {journal} {Class. Quant. Grav.}\ }\textbf {\bibinfo {volume} {39}},\ \bibinfo
  {pages} {105009} (\bibinfo {year} {2022})},\ \Eprint
  {https://arxiv.org/abs/2202.02355} {arXiv:2202.02355 [gr-qc]} \BibitemShut
  {NoStop}%
\bibitem [{\citenamefont {{Broderick}}\ \emph {et~al.}(2023)\citenamefont
  {{Broderick}}, \citenamefont {{Salehi}},\ and\ \citenamefont
  {{Georgiev}}}]{Broderick+2023}%
  \BibitemOpen
  \bibfield  {author} {\bibinfo {author} {\bibfnamefont {A.~E.}\ \bibnamefont
  {{Broderick}}}, \bibinfo {author} {\bibfnamefont {K.}~\bibnamefont
  {{Salehi}}},\ and\ \bibinfo {author} {\bibfnamefont {B.}~\bibnamefont
  {{Georgiev}}},\ }\bibfield  {title} {\bibinfo {title} {{Shadow Implications:
  What Does Measuring the Photon Ring Imply for Gravity?}},\ }\href
  {https://doi.org/10.3847/1538-4357/acf9f6} {\bibfield  {journal} {\bibinfo
  {journal} {\apj}\ }\textbf {\bibinfo {volume} {958}},\ \bibinfo {eid} {114}
  (\bibinfo {year} {2023})},\ \Eprint {https://arxiv.org/abs/2307.15120}
  {arXiv:2307.15120 [astro-ph.HE]} \BibitemShut {NoStop}%
\bibitem [{\citenamefont {Staelens}\ \emph {et~al.}(2023)\citenamefont
  {Staelens}, \citenamefont {Mayerson}, \citenamefont {Bacchini}, \citenamefont
  {Ripperda},\ and\ \citenamefont {K\"uchler}}]{Staelens+2023}%
  \BibitemOpen
  \bibfield  {author} {\bibinfo {author} {\bibfnamefont {S.}~\bibnamefont
  {Staelens}}, \bibinfo {author} {\bibfnamefont {D.~R.}\ \bibnamefont
  {Mayerson}}, \bibinfo {author} {\bibfnamefont {F.}~\bibnamefont {Bacchini}},
  \bibinfo {author} {\bibfnamefont {B.}~\bibnamefont {Ripperda}},\ and\
  \bibinfo {author} {\bibfnamefont {L.}~\bibnamefont {K\"uchler}},\ }\bibfield
  {title} {\bibinfo {title} {{Black hole photon rings beyond general
  relativity}},\ }\href {https://doi.org/10.1103/PhysRevD.107.124026}
  {\bibfield  {journal} {\bibinfo  {journal} {Phys. Rev. D}\ }\textbf {\bibinfo
  {volume} {107}},\ \bibinfo {pages} {124026} (\bibinfo {year} {2023})},\
  \Eprint {https://arxiv.org/abs/2303.02111} {arXiv:2303.02111 [gr-qc]}
  \BibitemShut {NoStop}%
\bibitem [{\citenamefont {Salehi}\ \emph {et~al.}(2023)\citenamefont {Salehi},
  \citenamefont {Broderick},\ and\ \citenamefont {Georgiev}}]{Salehi+2023}%
  \BibitemOpen
  \bibfield  {author} {\bibinfo {author} {\bibfnamefont {K.}~\bibnamefont
  {Salehi}}, \bibinfo {author} {\bibfnamefont {A.}~\bibnamefont {Broderick}},\
  and\ \bibinfo {author} {\bibfnamefont {B.}~\bibnamefont {Georgiev}},\
  }\bibfield  {title} {\bibinfo {title} {{Photon Rings and Shadow Size for
  General Integrable Spacetimes}},\ }\href@noop {} {\  (\bibinfo {year}
  {2023})},\ \Eprint {https://arxiv.org/abs/2311.01495} {arXiv:2311.01495
  [gr-qc]} \BibitemShut {NoStop}%
\bibitem [{\citenamefont {{Chael}}\ \emph {et~al.}(2021)\citenamefont
  {{Chael}}, \citenamefont {{Johnson}},\ and\ \citenamefont
  {{Lupsasca}}}]{Chael+2021}%
  \BibitemOpen
  \bibfield  {author} {\bibinfo {author} {\bibfnamefont {A.}~\bibnamefont
  {{Chael}}}, \bibinfo {author} {\bibfnamefont {M.~D.}\ \bibnamefont
  {{Johnson}}},\ and\ \bibinfo {author} {\bibfnamefont {A.}~\bibnamefont
  {{Lupsasca}}},\ }\bibfield  {title} {\bibinfo {title} {{Observing the Inner
  Shadow of a Black Hole: A Direct View of the Event Horizon}},\ }\href
  {https://doi.org/10.3847/1538-4357/ac09ee} {\bibfield  {journal} {\bibinfo
  {journal} {\apj}\ }\textbf {\bibinfo {volume} {918}},\ \bibinfo {eid} {6}
  (\bibinfo {year} {2021})},\ \Eprint {https://arxiv.org/abs/2106.00683}
  {arXiv:2106.00683 [astro-ph.HE]} \BibitemShut {NoStop}%
\bibitem [{\citenamefont {{Johnson}}\ \emph {et~al.}(2020)\citenamefont
  {{Johnson}} \emph {et~al.}}]{Johnson+2020}%
  \BibitemOpen
  \bibfield  {author} {\bibinfo {author} {\bibfnamefont {M.~D.}\ \bibnamefont
  {{Johnson}}} \emph {et~al.},\ }\bibfield  {title} {\bibinfo {title}
  {{Universal interferometric signatures of a black hole's photon ring}},\
  }\href {https://doi.org/10.1126/sciadv.aaz1310} {\bibfield  {journal}
  {\bibinfo  {journal} {\sciadv}\ }\textbf {\bibinfo {volume} {6}},\ \bibinfo
  {pages} {eaaz1310} (\bibinfo {year} {2020})},\ \Eprint
  {https://arxiv.org/abs/1907.04329} {arXiv:1907.04329 [astro-ph.IM]}
  \BibitemShut {NoStop}%
\bibitem [{\citenamefont {{Hadar}}\ \emph {et~al.}(2021)\citenamefont
  {{Hadar}}, \citenamefont {{Johnson}}, \citenamefont {{Lupsasca}},\ and\
  \citenamefont {{Wong}}}]{Hadar+2021}%
  \BibitemOpen
  \bibfield  {author} {\bibinfo {author} {\bibfnamefont {S.}~\bibnamefont
  {{Hadar}}}, \bibinfo {author} {\bibfnamefont {M.~D.}\ \bibnamefont
  {{Johnson}}}, \bibinfo {author} {\bibfnamefont {A.}~\bibnamefont
  {{Lupsasca}}},\ and\ \bibinfo {author} {\bibfnamefont {G.~N.}\ \bibnamefont
  {{Wong}}},\ }\bibfield  {title} {\bibinfo {title} {{Photon ring
  autocorrelations}},\ }\href {https://doi.org/10.1103/PhysRevD.103.104038}
  {\bibfield  {journal} {\bibinfo  {journal} {\prd}\ }\textbf {\bibinfo
  {volume} {103}},\ \bibinfo {eid} {104038} (\bibinfo {year} {2021})},\ \Eprint
  {https://arxiv.org/abs/2010.03683} {arXiv:2010.03683 [gr-qc]} \BibitemShut
  {NoStop}%
\bibitem [{\citenamefont {{Gralla}}\ and\ \citenamefont
  {{Lupsasca}}(2020)}]{Gralla+2020a}%
  \BibitemOpen
  \bibfield  {author} {\bibinfo {author} {\bibfnamefont {S.~E.}\ \bibnamefont
  {{Gralla}}}\ and\ \bibinfo {author} {\bibfnamefont {A.}~\bibnamefont
  {{Lupsasca}}},\ }\bibfield  {title} {\bibinfo {title} {{Lensing by Kerr black
  holes}},\ }\href {https://doi.org/10.1103/PhysRevD.101.044031} {\bibfield
  {journal} {\bibinfo  {journal} {\prd}\ }\textbf {\bibinfo {volume} {101}},\
  \bibinfo {eid} {044031} (\bibinfo {year} {2020})},\ \Eprint
  {https://arxiv.org/abs/1910.12873} {arXiv:1910.12873 [gr-qc]} \BibitemShut
  {NoStop}%
\bibitem [{\citenamefont {{Wong}}(2021)}]{Wong2021}%
  \BibitemOpen
  \bibfield  {author} {\bibinfo {author} {\bibfnamefont {G.~N.}\ \bibnamefont
  {{Wong}}},\ }\bibfield  {title} {\bibinfo {title} {{Black Hole Glimmer
  Signatures of Mass, Spin, and Inclination}},\ }\href
  {https://doi.org/10.3847/1538-4357/abdd2d} {\bibfield  {journal} {\bibinfo
  {journal} {\apj}\ }\textbf {\bibinfo {volume} {909}},\ \bibinfo {eid} {217}
  (\bibinfo {year} {2021})},\ \Eprint {https://arxiv.org/abs/2009.06641}
  {arXiv:2009.06641 [astro-ph.HE]} \BibitemShut {NoStop}%
\bibitem [{\citenamefont {{Ames}}\ and\ \citenamefont
  {{Thorne}}(1968)}]{Ames+1968}%
  \BibitemOpen
  \bibfield  {author} {\bibinfo {author} {\bibfnamefont {W.~L.}\ \bibnamefont
  {{Ames}}}\ and\ \bibinfo {author} {\bibfnamefont {K.~S.}\ \bibnamefont
  {{Thorne}}},\ }\bibfield  {title} {\bibinfo {title} {{The Optical Appearance
  of a Star that is Collapsing Through its Gravitational Radius}},\ }\href
  {https://doi.org/10.1086/149465} {\bibfield  {journal} {\bibinfo  {journal}
  {\apj}\ }\textbf {\bibinfo {volume} {151}},\ \bibinfo {pages} {659} (\bibinfo
  {year} {1968})}\BibitemShut {NoStop}%
\bibitem [{\citenamefont {{Cardoso}}\ \emph {et~al.}(2009)\citenamefont
  {{Cardoso}}, \citenamefont {{Miranda}}, \citenamefont {{Berti}},
  \citenamefont {{Witek}},\ and\ \citenamefont {{Zanchin}}}]{Cardoso+2009}%
  \BibitemOpen
  \bibfield  {author} {\bibinfo {author} {\bibfnamefont {V.}~\bibnamefont
  {{Cardoso}}}, \bibinfo {author} {\bibfnamefont {A.~S.}\ \bibnamefont
  {{Miranda}}}, \bibinfo {author} {\bibfnamefont {E.}~\bibnamefont {{Berti}}},
  \bibinfo {author} {\bibfnamefont {H.}~\bibnamefont {{Witek}}},\ and\ \bibinfo
  {author} {\bibfnamefont {V.~T.}\ \bibnamefont {{Zanchin}}},\ }\bibfield
  {title} {\bibinfo {title} {{Geodesic stability, Lyapunov exponents, and
  quasinormal modes}},\ }\href {https://doi.org/10.1103/PhysRevD.79.064016}
  {\bibfield  {journal} {\bibinfo  {journal} {\prd}\ }\textbf {\bibinfo
  {volume} {79}},\ \bibinfo {eid} {064016} (\bibinfo {year} {2009})},\ \Eprint
  {https://arxiv.org/abs/0812.1806} {arXiv:0812.1806 [hep-th]} \BibitemShut
  {NoStop}%
\bibitem [{\citenamefont {{Cardoso}}\ \emph {et~al.}(2021)\citenamefont
  {{Cardoso}}, \citenamefont {{Duque}},\ and\ \citenamefont
  {{Foschi}}}]{Cardoso+2021}%
  \BibitemOpen
  \bibfield  {author} {\bibinfo {author} {\bibfnamefont {V.}~\bibnamefont
  {{Cardoso}}}, \bibinfo {author} {\bibfnamefont {F.}~\bibnamefont {{Duque}}},\
  and\ \bibinfo {author} {\bibfnamefont {A.}~\bibnamefont {{Foschi}}},\
  }\bibfield  {title} {\bibinfo {title} {{Light ring and the appearance of
  matter accreted by black holes}},\ }\href
  {https://doi.org/10.1103/PhysRevD.103.104044} {\bibfield  {journal} {\bibinfo
   {journal} {\prd}\ }\textbf {\bibinfo {volume} {103}},\ \bibinfo {eid}
  {104044} (\bibinfo {year} {2021})},\ \Eprint
  {https://arxiv.org/abs/2102.07784} {arXiv:2102.07784 [gr-qc]} \BibitemShut
  {NoStop}%
\bibitem [{\citenamefont {{Paugnat}}\ \emph {et~al.}(2022)\citenamefont
  {{Paugnat}}, \citenamefont {{Lupsasca}}, \citenamefont {{Vincent}},\ and\
  \citenamefont {{Wielgus}}}]{Paugnat+2022}%
  \BibitemOpen
  \bibfield  {author} {\bibinfo {author} {\bibfnamefont {H.}~\bibnamefont
  {{Paugnat}}}, \bibinfo {author} {\bibfnamefont {A.}~\bibnamefont
  {{Lupsasca}}}, \bibinfo {author} {\bibfnamefont {F.~H.}\ \bibnamefont
  {{Vincent}}},\ and\ \bibinfo {author} {\bibfnamefont {M.}~\bibnamefont
  {{Wielgus}}},\ }\bibfield  {title} {\bibinfo {title} {{Photon ring test of
  the Kerr hypothesis: Variation in the ring shape}},\ }\href
  {https://doi.org/10.1051/0004-6361/202244216} {\bibfield  {journal} {\bibinfo
   {journal} {\aap}\ }\textbf {\bibinfo {volume} {668}},\ \bibinfo {eid} {A11}
  (\bibinfo {year} {2022})},\ \Eprint {https://arxiv.org/abs/2206.02781}
  {arXiv:2206.02781 [astro-ph.HE]} \BibitemShut {NoStop}%
\bibitem [{\citenamefont {{Vincent}}\ \emph {et~al.}(2022)\citenamefont
  {{Vincent}}, \citenamefont {{Gralla}}, \citenamefont {{Lupsasca}},\ and\
  \citenamefont {{Wielgus}}}]{Vincent+2022}%
  \BibitemOpen
  \bibfield  {author} {\bibinfo {author} {\bibfnamefont {F.~H.}\ \bibnamefont
  {{Vincent}}}, \bibinfo {author} {\bibfnamefont {S.~E.}\ \bibnamefont
  {{Gralla}}}, \bibinfo {author} {\bibfnamefont {A.}~\bibnamefont
  {{Lupsasca}}},\ and\ \bibinfo {author} {\bibfnamefont {M.}~\bibnamefont
  {{Wielgus}}},\ }\bibfield  {title} {\bibinfo {title} {{Images and photon ring
  signatures of thick disks around black holes}},\ }\href
  {https://doi.org/10.1051/0004-6361/202244339} {\bibfield  {journal} {\bibinfo
   {journal} {\aap}\ }\textbf {\bibinfo {volume} {667}},\ \bibinfo {eid} {A170}
  (\bibinfo {year} {2022})},\ \Eprint {https://arxiv.org/abs/2206.12066}
  {arXiv:2206.12066 [astro-ph.HE]} \BibitemShut {NoStop}%
\bibitem [{\citenamefont {{Kocherlakota}}\ \emph {et~al.}(prep)\citenamefont
  {{Kocherlakota}}, \citenamefont {{Rezzolla}}, \citenamefont {{Roy}},\ and\
  \citenamefont {{Wielgus}}}]{Kocherlakota+2024a}%
  \BibitemOpen
  \bibfield  {author} {\bibinfo {author} {\bibfnamefont {P.}~\bibnamefont
  {{Kocherlakota}}}, \bibinfo {author} {\bibfnamefont {L.}~\bibnamefont
  {{Rezzolla}}}, \bibinfo {author} {\bibfnamefont {R.}~\bibnamefont {{Roy}}},\
  and\ \bibinfo {author} {\bibfnamefont {M.}~\bibnamefont {{Wielgus}}},\
  }\href@noop {} {\  (\bibinfo {year} {in prep.})}\BibitemShut {NoStop}%
\bibitem [{\citenamefont {Hadar}\ \emph {et~al.}(2022)\citenamefont {Hadar},
  \citenamefont {Kapec}, \citenamefont {Lupsasca},\ and\ \citenamefont
  {Strominger}}]{Hadar+2022}%
  \BibitemOpen
  \bibfield  {author} {\bibinfo {author} {\bibfnamefont {S.}~\bibnamefont
  {Hadar}}, \bibinfo {author} {\bibfnamefont {D.}~\bibnamefont {Kapec}},
  \bibinfo {author} {\bibfnamefont {A.}~\bibnamefont {Lupsasca}},\ and\
  \bibinfo {author} {\bibfnamefont {A.}~\bibnamefont {Strominger}},\ }\bibfield
   {title} {\bibinfo {title} {{Holography of the photon ring}},\ }\href
  {https://doi.org/10.1088/1361-6382/ac8d43} {\bibfield  {journal} {\bibinfo
  {journal} {Class. Quant. Grav.}\ }\textbf {\bibinfo {volume} {39}},\ \bibinfo
  {pages} {215001} (\bibinfo {year} {2022})},\ \Eprint
  {https://arxiv.org/abs/2205.05064} {arXiv:2205.05064 [gr-qc]} \BibitemShut
  {NoStop}%
\bibitem [{\citenamefont {{Johannsen}}(2013)}]{Johannsen2013}%
  \BibitemOpen
  \bibfield  {author} {\bibinfo {author} {\bibfnamefont {T.}~\bibnamefont
  {{Johannsen}}},\ }\bibfield  {title} {\bibinfo {title} {{Systematic study of
  event horizons and pathologies of parametrically deformed Kerr spacetimes}},\
  }\href {https://doi.org/10.1103/PhysRevD.87.124017} {\bibfield  {journal}
  {\bibinfo  {journal} {\prd}\ }\textbf {\bibinfo {volume} {87}},\ \bibinfo
  {eid} {124017} (\bibinfo {year} {2013})},\ \Eprint
  {https://arxiv.org/abs/1304.7786} {arXiv:1304.7786 [gr-qc]} \BibitemShut
  {NoStop}%
\bibitem [{\citenamefont {{Rezzolla}}\ and\ \citenamefont
  {{Zhidenko}}(2014)}]{Rezzolla+2014}%
  \BibitemOpen
  \bibfield  {author} {\bibinfo {author} {\bibfnamefont {L.}~\bibnamefont
  {{Rezzolla}}}\ and\ \bibinfo {author} {\bibfnamefont {A.}~\bibnamefont
  {{Zhidenko}}},\ }\bibfield  {title} {\bibinfo {title} {{New parametrization
  for spherically symmetric black holes in metric theories of gravity}},\
  }\href {https://doi.org/10.1103/PhysRevD.90.084009} {\bibfield  {journal}
  {\bibinfo  {journal} {\prd}\ }\textbf {\bibinfo {volume} {90}},\ \bibinfo
  {eid} {084009} (\bibinfo {year} {2014})},\ \Eprint
  {https://arxiv.org/abs/1407.3086} {arXiv:1407.3086 [gr-qc]} \BibitemShut
  {NoStop}%
\bibitem [{\citenamefont {{Bauer}}\ \emph {et~al.}(2022)\citenamefont
  {{Bauer}}, \citenamefont {{C{\'a}rdenas-Avenda{\~n}o}}, \citenamefont
  {{Gammie}},\ and\ \citenamefont {{Yunes}}}]{Bauer+2022}%
  \BibitemOpen
  \bibfield  {author} {\bibinfo {author} {\bibfnamefont {A.~M.}\ \bibnamefont
  {{Bauer}}}, \bibinfo {author} {\bibfnamefont {A.}~\bibnamefont
  {{C{\'a}rdenas-Avenda{\~n}o}}}, \bibinfo {author} {\bibfnamefont {C.~F.}\
  \bibnamefont {{Gammie}}},\ and\ \bibinfo {author} {\bibfnamefont
  {N.}~\bibnamefont {{Yunes}}},\ }\bibfield  {title} {\bibinfo {title}
  {{Spherical Accretion in Alternative Theories of Gravity}},\ }\href
  {https://doi.org/10.3847/1538-4357/ac3a03} {\bibfield  {journal} {\bibinfo
  {journal} {\apj}\ }\textbf {\bibinfo {volume} {925}},\ \bibinfo {eid} {119}
  (\bibinfo {year} {2022})},\ \Eprint {https://arxiv.org/abs/2111.02178}
  {arXiv:2111.02178 [gr-qc]} \BibitemShut {NoStop}%
\bibitem [{\citenamefont {{Bozza}}\ and\ \citenamefont
  {{Scarpetta}}(2007)}]{Bozza+2007}%
  \BibitemOpen
  \bibfield  {author} {\bibinfo {author} {\bibfnamefont {V.}~\bibnamefont
  {{Bozza}}}\ and\ \bibinfo {author} {\bibfnamefont {G.}~\bibnamefont
  {{Scarpetta}}},\ }\bibfield  {title} {\bibinfo {title} {{Strong deflection
  limit of black hole gravitational lensing with arbitrary source distances}},\
  }\href {https://doi.org/10.1103/PhysRevD.76.083008} {\bibfield  {journal}
  {\bibinfo  {journal} {\prd}\ }\textbf {\bibinfo {volume} {76}},\ \bibinfo
  {eid} {083008} (\bibinfo {year} {2007})},\ \Eprint
  {https://arxiv.org/abs/0705.0246} {arXiv:0705.0246 [gr-qc]} \BibitemShut
  {NoStop}%
\bibitem [{\citenamefont {{Gralla}}\ \emph {et~al.}(2019)\citenamefont
  {{Gralla}}, \citenamefont {{Holz}},\ and\ \citenamefont
  {{Wald}}}]{Gralla+2019}%
  \BibitemOpen
  \bibfield  {author} {\bibinfo {author} {\bibfnamefont {S.~E.}\ \bibnamefont
  {{Gralla}}}, \bibinfo {author} {\bibfnamefont {D.~E.}\ \bibnamefont
  {{Holz}}},\ and\ \bibinfo {author} {\bibfnamefont {R.~M.}\ \bibnamefont
  {{Wald}}},\ }\bibfield  {title} {\bibinfo {title} {{Black hole shadows,
  photon rings, and lensing rings}},\ }\href
  {https://doi.org/10.1103/PhysRevD.100.024018} {\bibfield  {journal} {\bibinfo
   {journal} {\prd}\ }\textbf {\bibinfo {volume} {100}},\ \bibinfo {eid}
  {024018} (\bibinfo {year} {2019})},\ \Eprint
  {https://arxiv.org/abs/1906.00873} {arXiv:1906.00873 [astro-ph.HE]}
  \BibitemShut {NoStop}%
\bibitem [{\citenamefont {{Wielgus}}\ \emph {et~al.}(2020)\citenamefont
  {{Wielgus}}, \citenamefont {{Hor{\'a}k}}, \citenamefont {{Vincent}},\ and\
  \citenamefont {{Abramowicz}}}]{Wielgus+2020}%
  \BibitemOpen
  \bibfield  {author} {\bibinfo {author} {\bibfnamefont {M.}~\bibnamefont
  {{Wielgus}}}, \bibinfo {author} {\bibfnamefont {J.}~\bibnamefont
  {{Hor{\'a}k}}}, \bibinfo {author} {\bibfnamefont {F.}~\bibnamefont
  {{Vincent}}},\ and\ \bibinfo {author} {\bibfnamefont {M.}~\bibnamefont
  {{Abramowicz}}},\ }\bibfield  {title} {\bibinfo {title}
  {{Reflection-asymmetric wormholes and their double shadows}},\ }\href
  {https://doi.org/10.1103/PhysRevD.102.084044} {\bibfield  {journal} {\bibinfo
   {journal} {\prd}\ }\textbf {\bibinfo {volume} {102}},\ \bibinfo {eid}
  {084044} (\bibinfo {year} {2020})},\ \Eprint
  {https://arxiv.org/abs/2008.10130} {arXiv:2008.10130 [gr-qc]} \BibitemShut
  {NoStop}%
\bibitem [{\citenamefont {Gan}\ \emph {et~al.}(2021)\citenamefont {Gan},
  \citenamefont {Wang}, \citenamefont {Wu},\ and\ \citenamefont
  {Yang}}]{Gan+2021}%
  \BibitemOpen
  \bibfield  {author} {\bibinfo {author} {\bibfnamefont {Q.}~\bibnamefont
  {Gan}}, \bibinfo {author} {\bibfnamefont {P.}~\bibnamefont {Wang}}, \bibinfo
  {author} {\bibfnamefont {H.}~\bibnamefont {Wu}},\ and\ \bibinfo {author}
  {\bibfnamefont {H.}~\bibnamefont {Yang}},\ }\bibfield  {title} {\bibinfo
  {title} {Photon spheres and spherical accretion image of a hairy black
  hole},\ }\href {https://doi.org/10.1103/PhysRevD.104.024003} {\bibfield
  {journal} {\bibinfo  {journal} {Phys. Rev. D}\ }\textbf {\bibinfo {volume}
  {104}},\ \bibinfo {pages} {024003} (\bibinfo {year} {2021})}\BibitemShut
  {NoStop}%
\bibitem [{\citenamefont {{Guo}}\ \emph {et~al.}(2023)\citenamefont {{Guo}},
  \citenamefont {{Lu}}, \citenamefont {{Wang}}, \citenamefont {{Wu}},\ and\
  \citenamefont {{Yang}}}]{Guo+2022}%
  \BibitemOpen
  \bibfield  {author} {\bibinfo {author} {\bibfnamefont {G.}~\bibnamefont
  {{Guo}}}, \bibinfo {author} {\bibfnamefont {Y.}~\bibnamefont {{Lu}}},
  \bibinfo {author} {\bibfnamefont {P.}~\bibnamefont {{Wang}}}, \bibinfo
  {author} {\bibfnamefont {H.}~\bibnamefont {{Wu}}},\ and\ \bibinfo {author}
  {\bibfnamefont {H.}~\bibnamefont {{Yang}}},\ }\bibfield  {title} {\bibinfo
  {title} {{Black holes with multiple photon spheres}},\ }\href
  {https://doi.org/10.1103/PhysRevD.107.124037} {\bibfield  {journal} {\bibinfo
   {journal} {Phys. Rev. D}\ }\textbf {\bibinfo {volume} {107}},\ \bibinfo
  {pages} {124037} (\bibinfo {year} {2023})},\ \Eprint
  {https://arxiv.org/abs/2212.12901} {arXiv:2212.12901 [gr-qc]} \BibitemShut
  {NoStop}%
\bibitem [{\citenamefont {{Blaga}}\ \emph {et~al.}(2023)\citenamefont
  {{Blaga}}, \citenamefont {{Blaga}},\ and\ \citenamefont
  {{Harko}}}]{Blaga+2023}%
  \BibitemOpen
  \bibfield  {author} {\bibinfo {author} {\bibfnamefont {C.}~\bibnamefont
  {{Blaga}}}, \bibinfo {author} {\bibfnamefont {P.}~\bibnamefont {{Blaga}}},\
  and\ \bibinfo {author} {\bibfnamefont {T.}~\bibnamefont {{Harko}}},\
  }\bibfield  {title} {\bibinfo {title} {{Jacobi and Lyapunov stability
  analysis of circular geodesics around a spherically symmetric dilaton black
  hole}},\ }\href {https://doi.org/10.3390/sym15020329} {\bibfield  {journal}
  {\bibinfo  {journal} {Symmetry}\ }\textbf {\bibinfo {volume} {15}},\ \bibinfo
  {pages} {329} (\bibinfo {year} {2023})},\ \Eprint
  {https://arxiv.org/abs/2301.07678} {arXiv:2301.07678 [gr-qc]} \BibitemShut
  {NoStop}%
\bibitem [{\citenamefont {{Moriyama}}\ \emph {et~al.}(2019)\citenamefont
  {{Moriyama}}, \citenamefont {{Mineshige}}, \citenamefont {{Honma}},\ and\
  \citenamefont {{Akiyama}}}]{Moriyama+2019}%
  \BibitemOpen
  \bibfield  {author} {\bibinfo {author} {\bibfnamefont {K.}~\bibnamefont
  {{Moriyama}}}, \bibinfo {author} {\bibfnamefont {S.}~\bibnamefont
  {{Mineshige}}}, \bibinfo {author} {\bibfnamefont {M.}~\bibnamefont
  {{Honma}}},\ and\ \bibinfo {author} {\bibfnamefont {K.}~\bibnamefont
  {{Akiyama}}},\ }\bibfield  {title} {\bibinfo {title} {{Black Hole Spin
  Measurement Based on Time-domain VLBI Observations of Infalling Gas
  Clouds}},\ }\href {https://doi.org/10.3847/1538-4357/ab505b} {\bibfield
  {journal} {\bibinfo  {journal} {\apj}\ }\textbf {\bibinfo {volume} {887}},\
  \bibinfo {eid} {227} (\bibinfo {year} {2019})},\ \Eprint
  {https://arxiv.org/abs/1910.10713} {arXiv:1910.10713 [astro-ph.HE]}
  \BibitemShut {NoStop}%
\bibitem [{\citenamefont {{Wei}}(2020)}]{Wei2020}%
  \BibitemOpen
  \bibfield  {author} {\bibinfo {author} {\bibfnamefont {S.-W.}\ \bibnamefont
  {{Wei}}},\ }\bibfield  {title} {\bibinfo {title} {{Topological charge and
  black hole photon spheres}},\ }\href
  {https://doi.org/10.1103/PhysRevD.102.064039} {\bibfield  {journal} {\bibinfo
   {journal} {\prd}\ }\textbf {\bibinfo {volume} {102}},\ \bibinfo {eid}
  {064039} (\bibinfo {year} {2020})},\ \Eprint
  {https://arxiv.org/abs/2006.02112} {arXiv:2006.02112 [gr-qc]} \BibitemShut
  {NoStop}%
\bibitem [{\citenamefont {{Ye}}\ and\ \citenamefont {{Wei}}(2023)}]{Xu+2023}%
  \BibitemOpen
  \bibfield  {author} {\bibinfo {author} {\bibfnamefont {X.}~\bibnamefont
  {{Ye}}}\ and\ \bibinfo {author} {\bibfnamefont {S.-W.}\ \bibnamefont
  {{Wei}}},\ }\bibfield  {title} {\bibinfo {title} {{Distinct topological
  configurations of equatorial timelike circular orbit for spherically
  symmetric (hairy) black~holes}},\ }\href
  {https://doi.org/10.1088/1475-7516/2023/07/049} {\bibfield  {journal}
  {\bibinfo  {journal} {JCAP}\ }\textbf {\bibinfo {volume} {07}},\ \bibinfo
  {pages} {049}},\ \Eprint {https://arxiv.org/abs/2301.04786} {arXiv:2301.04786
  [gr-qc]} \BibitemShut {NoStop}%
\bibitem [{\citenamefont {{Virbhadra}}\ and\ \citenamefont
  {{Ellis}}(2000)}]{Virbhadra+2000}%
  \BibitemOpen
  \bibfield  {author} {\bibinfo {author} {\bibfnamefont {K.~S.}\ \bibnamefont
  {{Virbhadra}}}\ and\ \bibinfo {author} {\bibfnamefont {G.~F.~R.}\
  \bibnamefont {{Ellis}}},\ }\bibfield  {title} {\bibinfo {title}
  {{Schwarzschild black hole lensing}},\ }\href
  {https://doi.org/10.1103/PhysRevD.62.084003} {\bibfield  {journal} {\bibinfo
  {journal} {\prd}\ }\textbf {\bibinfo {volume} {62}},\ \bibinfo {eid} {084003}
  (\bibinfo {year} {2000})},\ \Eprint {https://arxiv.org/abs/astro-ph/9904193}
  {arXiv:astro-ph/9904193 [astro-ph]} \BibitemShut {NoStop}%
\bibitem [{\citenamefont {{Frittelli}}\ \emph {et~al.}(2000)\citenamefont
  {{Frittelli}}, \citenamefont {{Kling}},\ and\ \citenamefont
  {{Newman}}}]{Frittelli+2000}%
  \BibitemOpen
  \bibfield  {author} {\bibinfo {author} {\bibfnamefont {S.}~\bibnamefont
  {{Frittelli}}}, \bibinfo {author} {\bibfnamefont {T.~P.}\ \bibnamefont
  {{Kling}}},\ and\ \bibinfo {author} {\bibfnamefont {E.~T.}\ \bibnamefont
  {{Newman}}},\ }\bibfield  {title} {\bibinfo {title} {{Spacetime perspective
  of Schwarzschild lensing}},\ }\href
  {https://doi.org/10.1103/PhysRevD.61.064021} {\bibfield  {journal} {\bibinfo
  {journal} {\prd}\ }\textbf {\bibinfo {volume} {61}},\ \bibinfo {eid} {064021}
  (\bibinfo {year} {2000})},\ \Eprint {https://arxiv.org/abs/gr-qc/0001037}
  {arXiv:gr-qc/0001037 [gr-qc]} \BibitemShut {NoStop}%
\bibitem [{\citenamefont {{Narayan}}\ \emph {et~al.}(2022)\citenamefont
  {{Narayan}}, \citenamefont {{Chael}}, \citenamefont {{Chatterjee}},
  \citenamefont {{Ricarte}},\ and\ \citenamefont {{Curd}}}]{Narayan+2022}%
  \BibitemOpen
  \bibfield  {author} {\bibinfo {author} {\bibfnamefont {R.}~\bibnamefont
  {{Narayan}}}, \bibinfo {author} {\bibfnamefont {A.}~\bibnamefont {{Chael}}},
  \bibinfo {author} {\bibfnamefont {K.}~\bibnamefont {{Chatterjee}}}, \bibinfo
  {author} {\bibfnamefont {A.}~\bibnamefont {{Ricarte}}},\ and\ \bibinfo
  {author} {\bibfnamefont {B.}~\bibnamefont {{Curd}}},\ }\bibfield  {title}
  {\bibinfo {title} {{Jets in magnetically arrested hot accretion flows:
  geometry, power, and black hole spin-down}},\ }\href
  {https://doi.org/10.1093/mnras/stac285} {\bibfield  {journal} {\bibinfo
  {journal} {\mnras}\ }\textbf {\bibinfo {volume} {511}},\ \bibinfo {pages}
  {3795} (\bibinfo {year} {2022})},\ \Eprint {https://arxiv.org/abs/2108.12380}
  {arXiv:2108.12380 [astro-ph.HE]} \BibitemShut {NoStop}%
\bibitem [{\citenamefont {{Ohanian}}(1987)}]{Ohanian1987}%
  \BibitemOpen
  \bibfield  {author} {\bibinfo {author} {\bibfnamefont {H.~C.}\ \bibnamefont
  {{Ohanian}}},\ }\bibfield  {title} {\bibinfo {title} {{The black hole as a
  gravitational ``lens''}},\ }\href {https://doi.org/10.1119/1.15126}
  {\bibfield  {journal} {\bibinfo  {journal} {\ajp}\ }\textbf {\bibinfo
  {volume} {55}},\ \bibinfo {pages} {428} (\bibinfo {year} {1987})}\BibitemShut
  {NoStop}%
\bibitem [{\citenamefont {{Eckart}}\ \emph {et~al.}(2006)\citenamefont
  {{Eckart}} \emph {et~al.}}]{Eckart+2006}%
  \BibitemOpen
  \bibfield  {author} {\bibinfo {author} {\bibfnamefont {A.}~\bibnamefont
  {{Eckart}}} \emph {et~al.},\ }\bibfield  {title} {\bibinfo {title} {{The
  flare activity of Sagittarius A*. New coordinated mm to X-ray
  observations}},\ }\href {https://doi.org/10.1051/0004-6361:20054418}
  {\bibfield  {journal} {\bibinfo  {journal} {\aap}\ }\textbf {\bibinfo
  {volume} {450}},\ \bibinfo {pages} {535} (\bibinfo {year} {2006})},\ \Eprint
  {https://arxiv.org/abs/astro-ph/0512440} {arXiv:astro-ph/0512440 [astro-ph]}
  \BibitemShut {NoStop}%
\bibitem [{\citenamefont {{GRAVITY Collaboration}}\ \emph
  {et~al.}(2018)\citenamefont {{GRAVITY Collaboration}} \emph
  {et~al.}}]{Gravity+2018}%
  \BibitemOpen
  \bibfield  {author} {\bibinfo {author} {\bibnamefont {{GRAVITY
  Collaboration}}} \emph {et~al.},\ }\bibfield  {title} {\bibinfo {title}
  {{Detection of orbital motions near the last stable circular orbit of the
  massive black hole SgrA*}},\ }\href
  {https://doi.org/10.1051/0004-6361/201834294} {\bibfield  {journal} {\bibinfo
   {journal} {\aap}\ }\textbf {\bibinfo {volume} {618}},\ \bibinfo {eid} {L10}
  (\bibinfo {year} {2018})},\ \Eprint {https://arxiv.org/abs/1810.12641}
  {arXiv:1810.12641 [astro-ph.GA]} \BibitemShut {NoStop}%
\bibitem [{\citenamefont {{Tiede}}\ \emph {et~al.}(2020)\citenamefont {{Tiede}}
  \emph {et~al.}}]{Tiede+2020}%
  \BibitemOpen
  \bibfield  {author} {\bibinfo {author} {\bibfnamefont {P.}~\bibnamefont
  {{Tiede}}} \emph {et~al.},\ }\bibfield  {title} {\bibinfo {title} {{Spacetime
  Tomography Using the Event Horizon Telescope}},\ }\href
  {https://doi.org/10.3847/1538-4357/ab744c} {\bibfield  {journal} {\bibinfo
  {journal} {\apj}\ }\textbf {\bibinfo {volume} {892}},\ \bibinfo {eid} {132}
  (\bibinfo {year} {2020})},\ \Eprint {https://arxiv.org/abs/2002.05735}
  {arXiv:2002.05735 [astro-ph.HE]} \BibitemShut {NoStop}%
\bibitem [{\citenamefont {{Ball}}\ \emph {et~al.}(2021)\citenamefont {{Ball}},
  \citenamefont {{{\"O}zel}}, \citenamefont {{Christian}}, \citenamefont
  {{Chan}},\ and\ \citenamefont {{Psaltis}}}]{Ball+2021}%
  \BibitemOpen
  \bibfield  {author} {\bibinfo {author} {\bibfnamefont {D.}~\bibnamefont
  {{Ball}}}, \bibinfo {author} {\bibfnamefont {F.}~\bibnamefont {{{\"O}zel}}},
  \bibinfo {author} {\bibfnamefont {P.}~\bibnamefont {{Christian}}}, \bibinfo
  {author} {\bibfnamefont {C.-K.}\ \bibnamefont {{Chan}}},\ and\ \bibinfo
  {author} {\bibfnamefont {D.}~\bibnamefont {{Psaltis}}},\ }\bibfield  {title}
  {\bibinfo {title} {{A Plasmoid model for the Sgr A* Flares Observed With
  Gravity and CHANDRA}},\ }\href {https://doi.org/10.3847/1538-4357/abf8ae}
  {\bibfield  {journal} {\bibinfo  {journal} {\apj}\ }\textbf {\bibinfo
  {volume} {917}},\ \bibinfo {eid} {8} (\bibinfo {year} {2021})},\ \Eprint
  {https://arxiv.org/abs/2005.14251} {arXiv:2005.14251 [astro-ph.HE]}
  \BibitemShut {NoStop}%
\bibitem [{\citenamefont {{Sahu}}\ \emph {et~al.}(2013)\citenamefont {{Sahu}},
  \citenamefont {{Patil}}, \citenamefont {{Narasimha}},\ and\ \citenamefont
  {{Joshi}}}]{Sahu+2013}%
  \BibitemOpen
  \bibfield  {author} {\bibinfo {author} {\bibfnamefont {S.}~\bibnamefont
  {{Sahu}}}, \bibinfo {author} {\bibfnamefont {M.}~\bibnamefont {{Patil}}},
  \bibinfo {author} {\bibfnamefont {D.}~\bibnamefont {{Narasimha}}},\ and\
  \bibinfo {author} {\bibfnamefont {P.~S.}\ \bibnamefont {{Joshi}}},\
  }\bibfield  {title} {\bibinfo {title} {{Time delay between relativistic
  images as a probe of cosmic censorship}},\ }\href
  {https://doi.org/10.1103/PhysRevD.88.103002} {\bibfield  {journal} {\bibinfo
  {journal} {\prd}\ }\textbf {\bibinfo {volume} {88}},\ \bibinfo {eid} {103002}
  (\bibinfo {year} {2013})},\ \Eprint {https://arxiv.org/abs/1310.5350}
  {arXiv:1310.5350 [gr-qc]} \BibitemShut {NoStop}%
\bibitem [{\citenamefont {Chen}\ \emph
  {et~al.}(2023{\natexlab{a}})\citenamefont {Chen}, \citenamefont {Xue},
  \citenamefont {Brito},\ and\ \citenamefont {Cardoso}}]{Chen+2023b}%
  \BibitemOpen
  \bibfield  {author} {\bibinfo {author} {\bibfnamefont {Y.}~\bibnamefont
  {Chen}}, \bibinfo {author} {\bibfnamefont {X.}~\bibnamefont {Xue}}, \bibinfo
  {author} {\bibfnamefont {R.}~\bibnamefont {Brito}},\ and\ \bibinfo {author}
  {\bibfnamefont {V.}~\bibnamefont {Cardoso}},\ }\bibfield  {title} {\bibinfo
  {title} {{Photon Ring Astrometry for Superradiant Clouds}},\ }\href
  {https://doi.org/10.1103/PhysRevLett.130.111401} {\bibfield  {journal}
  {\bibinfo  {journal} {Phys. Rev. Lett.}\ }\textbf {\bibinfo {volume} {130}},\
  \bibinfo {pages} {111401} (\bibinfo {year} {2023}{\natexlab{a}})},\ \Eprint
  {https://arxiv.org/abs/2211.03794} {arXiv:2211.03794 [gr-qc]} \BibitemShut
  {NoStop}%
\bibitem [{\citenamefont {{Stefanov}}\ \emph {et~al.}(2010)\citenamefont
  {{Stefanov}}, \citenamefont {{Yazadjiev}},\ and\ \citenamefont
  {{Gyulchev}}}]{Stefanov+2010}%
  \BibitemOpen
  \bibfield  {author} {\bibinfo {author} {\bibfnamefont {I.~Z.}\ \bibnamefont
  {{Stefanov}}}, \bibinfo {author} {\bibfnamefont {S.~S.}\ \bibnamefont
  {{Yazadjiev}}},\ and\ \bibinfo {author} {\bibfnamefont {G.~G.}\ \bibnamefont
  {{Gyulchev}}},\ }\bibfield  {title} {\bibinfo {title} {{Connection between
  Black-Hole Quasinormal Modes and Lensing in the Strong Deflection Limit}},\
  }\href {https://doi.org/10.1103/PhysRevLett.104.251103} {\bibfield  {journal}
  {\bibinfo  {journal} {\prl}\ }\textbf {\bibinfo {volume} {104}},\ \bibinfo
  {eid} {251103} (\bibinfo {year} {2010})},\ \Eprint
  {https://arxiv.org/abs/1003.1609} {arXiv:1003.1609 [gr-qc]} \BibitemShut
  {NoStop}%
\bibitem [{\citenamefont {Chen}\ \emph
  {et~al.}(2023{\natexlab{b}})\citenamefont {Chen}, \citenamefont {Chen},
  \citenamefont {Ho},\ and\ \citenamefont {Tseng}}]{Chen+2023a}%
  \BibitemOpen
  \bibfield  {author} {\bibinfo {author} {\bibfnamefont {C.-Y.}\ \bibnamefont
  {Chen}}, \bibinfo {author} {\bibfnamefont {Y.-J.}\ \bibnamefont {Chen}},
  \bibinfo {author} {\bibfnamefont {M.-Y.}\ \bibnamefont {Ho}},\ and\ \bibinfo
  {author} {\bibfnamefont {Y.-H.}\ \bibnamefont {Tseng}},\ }\bibfield  {title}
  {\bibinfo {title} {{A novel test of gravity via black hole eikonal
  correspondence}},\ }\href {https://doi.org/10.1016/j.physletb.2023.138153}
  {\bibfield  {journal} {\bibinfo  {journal} {Phys. Lett. B}\ }\textbf
  {\bibinfo {volume} {845}},\ \bibinfo {pages} {138153} (\bibinfo {year}
  {2023}{\natexlab{b}})},\ \Eprint {https://arxiv.org/abs/2212.10028}
  {arXiv:2212.10028 [gr-qc]} \BibitemShut {NoStop}%
\bibitem [{\citenamefont {Geroch}(1970)}]{Geroch1970}%
  \BibitemOpen
  \bibfield  {author} {\bibinfo {author} {\bibfnamefont {R.~P.}\ \bibnamefont
  {Geroch}},\ }\bibfield  {title} {\bibinfo {title} {{Multipole moments. II.
  Curved space}},\ }\href {https://doi.org/10.1063/1.1665427} {\bibfield
  {journal} {\bibinfo  {journal} {J. Math. Phys.}\ }\textbf {\bibinfo {volume}
  {11}},\ \bibinfo {pages} {2580} (\bibinfo {year} {1970})}\BibitemShut
  {NoStop}%
\bibitem [{\citenamefont {{Vigeland}}\ \emph {et~al.}(2011)\citenamefont
  {{Vigeland}}, \citenamefont {{Yunes}},\ and\ \citenamefont
  {{Stein}}}]{Vigeland+2011}%
  \BibitemOpen
  \bibfield  {author} {\bibinfo {author} {\bibfnamefont {S.}~\bibnamefont
  {{Vigeland}}}, \bibinfo {author} {\bibfnamefont {N.}~\bibnamefont
  {{Yunes}}},\ and\ \bibinfo {author} {\bibfnamefont {L.~C.}\ \bibnamefont
  {{Stein}}},\ }\bibfield  {title} {\bibinfo {title} {{Bumpy black holes in
  alternative theories of gravity}},\ }\href
  {https://doi.org/10.1103/PhysRevD.83.104027} {\bibfield  {journal} {\bibinfo
  {journal} {\prd}\ }\textbf {\bibinfo {volume} {83}},\ \bibinfo {eid} {104027}
  (\bibinfo {year} {2011})},\ \Eprint {https://arxiv.org/abs/1102.3706}
  {arXiv:1102.3706 [gr-qc]} \BibitemShut {NoStop}%
\bibitem [{\citenamefont {{Kocherlakota}}\ and\ \citenamefont
  {{Rezzolla}}(2020)}]{Kocherlakota+2020}%
  \BibitemOpen
  \bibfield  {author} {\bibinfo {author} {\bibfnamefont {P.}~\bibnamefont
  {{Kocherlakota}}}\ and\ \bibinfo {author} {\bibfnamefont {L.}~\bibnamefont
  {{Rezzolla}}},\ }\bibfield  {title} {\bibinfo {title} {{Accurate mapping of
  spherically symmetric black holes in a parametrized framework}},\ }\href
  {https://doi.org/10.1103/PhysRevD.102.064058} {\bibfield  {journal} {\bibinfo
   {journal} {\prd}\ }\textbf {\bibinfo {volume} {102}},\ \bibinfo {eid}
  {064058} (\bibinfo {year} {2020})},\ \Eprint
  {https://arxiv.org/abs/2007.15593} {arXiv:2007.15593 [gr-qc]} \BibitemShut
  {NoStop}%
\bibitem [{\citenamefont {{Konoplya}}\ and\ \citenamefont
  {{Zhidenko}}(2020)}]{Konoplya+2020}%
  \BibitemOpen
  \bibfield  {author} {\bibinfo {author} {\bibfnamefont {R.~A.}\ \bibnamefont
  {{Konoplya}}}\ and\ \bibinfo {author} {\bibfnamefont {A.}~\bibnamefont
  {{Zhidenko}}},\ }\bibfield  {title} {\bibinfo {title} {{General
  parametrization of black holes: The only parameters that matter}},\ }\href
  {https://doi.org/10.1103/PhysRevD.101.124004} {\bibfield  {journal} {\bibinfo
   {journal} {\prd}\ }\textbf {\bibinfo {volume} {101}},\ \bibinfo {eid}
  {124004} (\bibinfo {year} {2020})},\ \Eprint
  {https://arxiv.org/abs/2001.06100} {arXiv:2001.06100 [gr-qc]} \BibitemShut
  {NoStop}%
\bibitem [{\citenamefont {{Arnowitt}}\ \emph {et~al.}(2008)\citenamefont
  {{Arnowitt}}, \citenamefont {{Deser}},\ and\ \citenamefont
  {{Misner}}}]{Arnowitt+2008}%
  \BibitemOpen
  \bibfield  {author} {\bibinfo {author} {\bibfnamefont {R.}~\bibnamefont
  {{Arnowitt}}}, \bibinfo {author} {\bibfnamefont {S.}~\bibnamefont
  {{Deser}}},\ and\ \bibinfo {author} {\bibfnamefont {C.~W.}\ \bibnamefont
  {{Misner}}},\ }\bibfield  {title} {\bibinfo {title} {{Republication of: The
  dynamics of general relativity}},\ }\href
  {https://doi.org/10.1007/s10714-008-0661-1} {\bibfield  {journal} {\bibinfo
  {journal} {General Relativity and Gravitation}\ }\textbf {\bibinfo {volume}
  {40}},\ \bibinfo {pages} {1997} (\bibinfo {year} {2008})},\ \Eprint
  {https://arxiv.org/abs/gr-qc/0405109} {arXiv:gr-qc/0405109 [gr-qc]}
  \BibitemShut {NoStop}%
\bibitem [{\citenamefont {{Will}}(2014)}]{Will2014}%
  \BibitemOpen
  \bibfield  {author} {\bibinfo {author} {\bibfnamefont {C.~M.}\ \bibnamefont
  {{Will}}},\ }\bibfield  {title} {\bibinfo {title} {{The Confrontation between
  General Relativity and Experiment}},\ }\href
  {https://doi.org/10.12942/lrr-2014-4} {\bibfield  {journal} {\bibinfo
  {journal} {Living Reviews in Relativity}\ }\textbf {\bibinfo {volume} {17}},\
  \bibinfo {eid} {4} (\bibinfo {year} {2014})},\ \Eprint
  {https://arxiv.org/abs/1403.7377} {arXiv:1403.7377 [gr-qc]} \BibitemShut
  {NoStop}%
\bibitem [{\citenamefont {{Wald}}(1984)}]{Wald1984}%
  \BibitemOpen
  \bibfield  {author} {\bibinfo {author} {\bibfnamefont {R.~M.}\ \bibnamefont
  {{Wald}}},\ }\href@noop {} {\emph {\bibinfo {title} {{General Relativity}}}}\
  (\bibinfo  {publisher} {U. Chicago Press},\ \bibinfo {year}
  {1984})\BibitemShut {NoStop}%
\bibitem [{\citenamefont {Akiyama}\ \emph
  {et~al.}(2022{\natexlab{d}})\citenamefont {Akiyama} \emph
  {et~al.}}]{EHTC+2022c}%
  \BibitemOpen
  \bibfield  {author} {\bibinfo {author} {\bibfnamefont {K.}~\bibnamefont
  {Akiyama}} \emph {et~al.} (\bibinfo {collaboration} {Event Horizon
  Telescope}),\ }\bibfield  {title} {\bibinfo {title} {{First Sagittarius A*
  Event Horizon Telescope Results. III. Imaging of the Galactic Center
  Supermassive Black Hole}},\ }\href {https://doi.org/10.3847/2041-8213/ac6429}
  {\bibfield  {journal} {\bibinfo  {journal} {Astrophys. J. Lett.}\ }\textbf
  {\bibinfo {volume} {930}},\ \bibinfo {pages} {L14} (\bibinfo {year}
  {2022}{\natexlab{d}})},\ \Eprint {https://arxiv.org/abs/2311.09479}
  {arXiv:2311.09479 [astro-ph.HE]} \BibitemShut {NoStop}%
\bibitem [{\citenamefont {Porth}\ \emph {et~al.}(2019)\citenamefont {Porth}
  \emph {et~al.}}]{Porth+2019}%
  \BibitemOpen
  \bibfield  {author} {\bibinfo {author} {\bibfnamefont {O.}~\bibnamefont
  {Porth}} \emph {et~al.} (\bibinfo {collaboration} {Event Horizon
  Telescope}),\ }\bibfield  {title} {\bibinfo {title} {{The Event Horizon
  General Relativistic Magnetohydrodynamic Code Comparison Project}},\ }\href
  {https://doi.org/10.3847/1538-4365/ab29fd} {\bibfield  {journal} {\bibinfo
  {journal} {Astrophys. J. Suppl.}\ }\textbf {\bibinfo {volume} {243}},\
  \bibinfo {pages} {26} (\bibinfo {year} {2019})},\ \Eprint
  {https://arxiv.org/abs/1904.04923} {arXiv:1904.04923 [astro-ph.HE]}
  \BibitemShut {NoStop}%
\bibitem [{\citenamefont {{Chatterjee}}\ and\ \citenamefont
  {{Narayan}}(2022)}]{Chatterjee+2022}%
  \BibitemOpen
  \bibfield  {author} {\bibinfo {author} {\bibfnamefont {K.}~\bibnamefont
  {{Chatterjee}}}\ and\ \bibinfo {author} {\bibfnamefont {R.}~\bibnamefont
  {{Narayan}}},\ }\bibfield  {title} {\bibinfo {title} {{Flux Eruption Events
  Drive Angular Momentum Transport in Magnetically Arrested Accretion Flows}},\
  }\href {https://doi.org/10.3847/1538-4357/ac9d97} {\bibfield  {journal}
  {\bibinfo  {journal} {\apj}\ }\textbf {\bibinfo {volume} {941}},\ \bibinfo
  {eid} {30} (\bibinfo {year} {2022})},\ \Eprint
  {https://arxiv.org/abs/2210.08045} {arXiv:2210.08045 [astro-ph.HE]}
  \BibitemShut {NoStop}%
\bibitem [{\citenamefont {{Ayzenberg}}\ \emph {et~al.}(2023)\citenamefont
  {{Ayzenberg}} \emph {et~al.}}]{Ayzenberg+2023}%
  \BibitemOpen
  \bibfield  {author} {\bibinfo {author} {\bibfnamefont {D.}~\bibnamefont
  {{Ayzenberg}}} \emph {et~al.},\ }\bibfield  {title} {\bibinfo {title}
  {{Fundamental Physics Opportunities with the Next-Generation Event Horizon
  Telescope}},\ }\href@noop {} {\bibfield  {journal} {\bibinfo  {journal}
  {arXiv}\ ,\ \bibinfo {eid} {arXiv:2312.02130}} (\bibinfo {year} {2023})},\
  \Eprint {https://arxiv.org/abs/2312.02130} {arXiv:2312.02130 [astro-ph.HE]}
  \BibitemShut {NoStop}%
\end{thebibliography}%

\begin{appendix}

\begin{figure*}
\centering
\includegraphics[width=1.9\columnwidth]{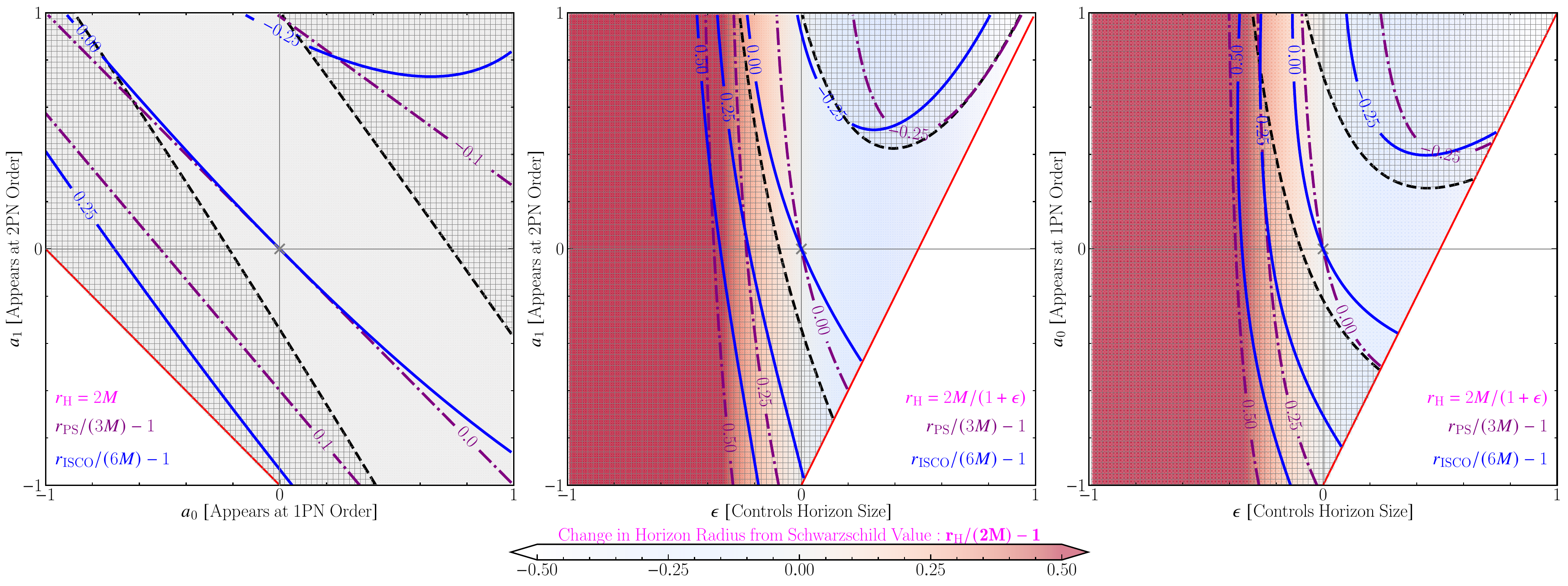}
\caption{\textit{Variation in characteristic spacetime locations with metric deviation parameters.} 
We show the variation in the locations of the horizon ($r_{\mathrm{H}}$; colormap), the photon sphere ($r_{\mathrm{PS}}$; dot-dashed lines), and the innermost stable circular orbit ($r_{\mathrm{ISCO}}$; solid lines) for three different families of Rezzolla-Zhidenko BHs across three different panels. All of these BHs have the same mass $M$, and the Schwarzschild BH is located at $(0, 0)$ in each panel. The hatched regions are disallowed by the 2017 EHT ($1\sigma$) shadow size measurement of Sgr A$^\star$. This shows how the EHT measurements can be translated into $\approx 25\%$ constraints on the deviations of these characteristic locations, for spherically-symmetric black holes, from their Schwarzschild values.}
\label{fig:FigA1_RZ_Pure_Metric_Characteristics}
\end{figure*}

\section{The Horizon, Photon Sphere, and ISCO in RZ BHs}

We now show Fig. \ref{fig:FigA1_RZ_Pure_Metric_Characteristics} the impact of varying the spacetime geometry on the locations of the event horizon, the photon sphere, and the innermost stable circular orbit (ISCO; This is the timelike Keplerian geodesic that is closest to the BH), which are determined purely by the spacetime metric. In particular, we consider Rezzolla-Zhidenko BHs that are described by the deviation parameters $\epsilon, a_0$, and $ a_1$. Since, as discussed above (see also Ref. \cite{Kocherlakota+2020}), the locations of the horizon, the photon sphere, and the ISCO are set by the $tt-$component of the metric alone, in areal-polar coordinates ($R(r) = r$), varying $b_0$ or $b_1$ has no effect here. 

While these characteristic spacetime features are not directly observable, they play an important role in shaping our understanding of the physics of black holes. In each panel, we only vary, between $\pm 1$, the metric deviation parameters that are shown on the axes (all others are set to zero). It is easy to see that the most sensitive variations occur due to changes in $\epsilon$, i.e., due to changes in the horizon size. The hatched regions correspond to those that are ruled out by recent EHT observations, as discussed above.

\end{appendix}

\end{document}